\documentclass[12pt]{article}
\pdfoutput=1
\usepackage{putex}
\usepackage{graphicx}
\usepackage{epstopdf}
\usepackage{enumerate}
\usepackage{cite}
\usepackage{tensor}
\usepackage{slashed}
\usepackage{feynmf}
\usepackage{hyperref}
\usepackage{float}
\usepackage[bottom]{footmisc}

\newcommand{\normord}[1]{:\mathrel{#1}:}

\usepackage[usenames,dvipsnames,svgnames,table]{xcolor}

\hypersetup{
	pdfstartview={FitH},    
	pdftitle={Quantum Thermalization and Virasoro Symmetry},    
	pdfauthor={Besken, Datta, Kraus},     
	colorlinks=true,       
	linkcolor=blue,          
	citecolor=blue!59!black! ,        
	filecolor=magenta,      
	urlcolor=blue!75!black,           
	linktoc=all
}

\makeatletter
\g@addto@macro\bfseries{\boldmath}
\makeatother

\numberwithin{equation}{section}

\def\cK{{\cal K}}

\def\d{\delta}

\def\Oc{{\cal O}}

\def\Lc{{\cal L}}

\newcommand {\be} {\begin {equation}}
\newcommand {\ee} {\end {equation}}
\newcommand {\nn} {\nonumber}
\newcommand {\bes} {\begin {equation*}}
\newcommand {\ees} {\end {equation*}}

\newcommand{\R}{\mathbb{R}}
\newcommand{\C}{\mathbb{C}}

\newcommand{\beq}{\begin{equation}}
\newcommand{\eeq}{\end{equation}}

\newcommand{\p}{\partial}

\def\be{ \begin{equation} }
\def\ee{ \end{equation} }

\def\C{{\cal C}}

\def\C{{\cal C}}

\def\Tr{\mathop{\rm Tr}}

\newcommand{\bea}{\begin{eqnarray}}
\newcommand{\eea}{\end{eqnarray}}

\newcommand\ov{\over}

\def\C{\mathcal{C}}

\def\ub{\overline{u}}
\def\vb{\overline{v}}

\def\rt{\rightarrow}

\def\lb{{\overline{l}}}
\def\Ub{{\overline{U}}}
\def\Vb{{\overline{V}}}
\def\ov{\overline}

\def\vev#1{\langle #1 \rangle}
\def\pd{\partial}

\newcommand{\<}{\langle}
\renewcommand{\>}{\rangle}

\begin{document}

	\institution{UCLA}{ \quad\quad\quad\quad\quad\quad\quad\quad\ \,  Mani L. Bhaumik Institute for Theoretical Physics
		\cr Department of Physics \& Astronomy,\,University of California,\,Los Angeles,\,CA\,90095,\,USA}

	\title{Quantum Thermalization \\ and Virasoro Symmetry
	}
	
	\authors{Mert Be\c sken, Shouvik Datta and Per Kraus}
	
	\abstract{We initiate a systematic study of high energy matrix elements of local operators in 2d CFT.  Knowledge of these is required in order to determine whether the eigenstate thermalization hypothesis (ETH) can hold in such theories.  Most high energy states are high level Virasoro descendants, and by employing an oscillator representation of the Virasoro algebra we develop an efficient method for computing matrix elements of primary operators in such states.  In parameter regimes where we expect (e.g.~from AdS/CFT intuition)  thermalization to occur, we observe striking patterns in the matrix elements: diagonal matrix elements are smoothly varying and off-diagonal elements, while nonzero, are power-law suppressed compared to the diagonal elements. We discuss the implications of these universal properties of 2d CFTs in regard to their compatibility with ETH.
 }
	
	\date{}
	
	\maketitle
	\setcounter{tocdepth}{2}
	\begingroup
	\hypersetup{linkcolor=black}
	\tableofcontents
	\endgroup
	

\section{Introduction}


 In this paper we study matrix elements of local operators in two-dimensional conformal field theories (CFTs),
\bea
O_{ab} = \langle \psi_a | O|\psi_b\rangle~.
\eea
The states $|\psi_a\rangle$ are energy eigenstates.   Our motivations are twofold: first to explore some universal aspects of CFTs which have so far escaped attention, and second to address the question of whether such theories can exhibit thermalization and be  compatible with the (generalized) eigenstate thermalization hypothesis (ETH)  \cite{1991PhRvA..43.2046D,Srednicki:1994}.

We begin with some general comments about thermalization and ETH in generic quantum systems with many degrees of freedom; reviews include    \cite{2016AdPhy..65..239D,deutsch2018eigenstate}.   Questions regarding thermalization have to do with comparing quantities (meaning expectation values) evaluated in particular microstates to those computed in a thermal ensemble, which is here taken to be governed by the Boltzmann density matrix, $\rho = {1\over Z} e^{-\beta H}$.  If for a given observable $O$ we have $\langle \psi|O |\psi\rangle \approx \Tr(\rho O)$ then we say that the system has ``thermalized" with respect to that observable.

A necessary, but far from sufficient,  condition underlying the usefulness of a statistical description is that for all ``physically accessible" observables  ``almost all" states are approximately thermal.\footnote{For more on the meaning of ``almost all" we direct the reader to the references cited above, and here simply note that it rules out states such as those corresponding to a quantum superposition of states with macroscopically distinct properties. For example --- and this will be relevant for our discussion --- if the system possesses a conserved charge  $Q$ we may wish to rule out superpositions of states with macroscopically different values of $Q$.}  We refer to this condition as ``typicality";  its widespread occurrence holds by virtue of the central limit theorem.   Namely, if we consider an observable such as the total energy within some region, it can be written as the sum of a large number of weakly correlated contributions, and hence is sharply peaked over the ensemble of states.  The upshot is that if we prepare a quantum system in a typical state we expect that most observables will have thermal values.

As noted, typicality is a very weak condition, much weaker than is needed to establish thermalization.  For instance, even a free field theory exhibits typicality but clearly does not thermalize to the canonical ensemble.   Thermalization is the process by which an initial atypical and non-thermal state evolves under Hamiltonian time evolution to a thermal state.  Needless to say, most of experimental science, not to mention the existence of scientists, relies on  such  non-thermal states.  In a thermalizing system,  operator expectation values $\langle \psi|O(t)|\psi\rangle$ start out far from equilibrium and then eventually undergo small fluctuations around their thermal values for almost all times, with sporadic large fluctuations.    The demand is that this behavior be exhibited for all non-equilibrium initial states that can be prepared via a physically reasonable process.  An example of such an initial state consists of a collection of atoms confined to a small region by a partition which is subsequently removed. As before, we may treat as unphysical those initial states which involve superpositions of states of macroscopically different values of a certain conserved charge.

The ETH translates these physically intuitive but mathematically imprecise notions regarding thermalization into statements about matrix elements of operators in energy eigenstates.   	 The ETH ansatz for matrix elements is \cite{Srednicki:1994}
\bea\label{eth}
\langle \psi_a|O|\psi_b\rangle \approx \overline{O} \delta_{ab} + e^{-S\left({E_a+E_b\over 2}\right)/2} f_{O}(E_a,E_b) R_{ab}~.
\eea
Here the energies $E_a$ and $E_b$ are taken to be nearby in the sense that $E_a-E_b$ is much less than the total energy of the system, but we still allow for there to be many energies between these levels.  $\overline{O}$ is the thermal average computed at a temperature chosen such that $\overline{H}\approx E_{a,b}$, or alternatively is the microcanonical average computed from a narrow band of energy eigenstates including $E_{a,b}$.  $S$ is the entropy computed from the ensembles just mentioned; $f(E_a,E_b)$ is a slowly varying function; and $R_{ab}$ is a random matrix of zero mean and unit variance.   This form is motivated by consideration of expectation values in  states  $|\psi(t)\rangle = \sum_a c_ae^{-iE_a t} |\psi_a\rangle$, where the sum runs over states in some narrow energy window.   An out-of-equilibrium initial state can be created by choosing the $c_a$ such that the off-diagonal terms compete with the diagonal terms.  As time increases dephasing occurs and the off-diagonal elements approximately cancel out.  One is left with only the diagonal terms, and the form $O_{aa} = \overline{O}$ ensures that the thermal value results for any choice of $c_a$.  The  ETH form of the off-diagonal elements is motivated by various arguments and is supported by exact diagonalization studies of chaotic quantum systems \cite{rigol2008thermalization,huse2014,2016PhRvE..93c2104M,2017PhRvE..96a2157M}.    A key fact that distinguishes chaotic systems from integrable systems is that for the latter we expect most of the off-diagonal matrix elements to vanish due to selection rules, while for a chaotic systems we expect essentially all of them to be nonzero but small (e.g.~\cite{khatami2013fluctuation}).

The discussion presented so far needs refinement if the system possesses a physically relevant conserved charge $Q$.  By a ``physically relevant" charge we mean one for which it is possible, by a physically reasonable process,  to produce states in which the charge takes a value which is sharply defined yet different from its value in the canonical ensemble at the corresponding temperature.  In such a case we should ask about thermalization to the Gibbs ensemble in which a chemical potential for the conserved charge is included.  We should also refine the ETH ansatz to only include matrix elements of states with nearby (i.e. macroscopically indistinguishable) values of the charge.  We then speak of a ``generalized eigenstate thermalization hypothesis" \cite{Cassidy,Essler:2016ufo,VidmarRigol}. 


An extreme example is the case of free theories (quadratic Hamiltonians). Here there are an infinite number of conserved charges, which we can think of as the occupation numbers of individual particle states.   Such theories have been studied in this context in \cite{Sotiriadis:2014uza,Sotiriadis:2015kfa,Sotiriadis:2015xia,Mandal:2015kxi,Bastianello:2016nut}.    A simple example is a quantum quench in which a free field theory is prepared in the ground state of the theory with one value of the mass, and then allowed to evolve according to the Hamiltonian with a different mass.  Thermalization to the generalized Gibbs ensemble (GGE) is found to occur. 
The occurrence of thermalization in this and related cases is easily understood along the following lines.   First consider the case in which the initial wavefunction is Gaussian, as in the quench example just mentioned.   In such a state, correlation functions of local operators are fully determined by two-point functions.   However, it's easy to see that the (time averaged) two-point functions are fixed by the values of the conserved charges.  This implies thermalization to the GGE. 

What if the initial state is non-Gaussian?  Not any such state is allowed, rather we should demand a clustering property: connected correlators should fall off at least as a power in the separation.  This is required in order that the conserved charges take sharply defined values (for example, it rules out macroscopic superposition of Schrodinger cat type). In \cite{Murthy:2019tzf,bastianello2017quasi,gluza2019equilibration} it was argued that such states ``Gaussify", meaning that the wavefunction effectively becomes Gaussian at sufficiently late times.  The discussion of the last paragraph then applies.    

Much less is understood about interacting theories.  A special but  important class of such theories are 2d CFTs.   Previous studies include \cite{Asplund:2014coa,Lashkari:2016vgj,deBoer:2016bov,Basu:2017kzo,He:2017txy,He:2017vyf,Chen:2017yze,Faulkner:2017hll,Brehm:2018ipf,Romero-Bermudez:2018dim,Hikida:2018khg,Lashkari:2017hwq,Anous:2019yku,lai2015entanglement,Datta:2019jeo,Guo:2018pvi,Maloney:2018hdg,Maloney:2018yrz,Dymarsky:2018lhf,Dymarsky:2019etq,Dymarsky:2018iwx,Brehm:2019fyy,Sabella-Garnier:2019tsi}.  On the one hand, 2d CFTs have an infinite number of mutually commuting conserved KdV charges associated with  the conformal symmetry \cite{Bazhanov:1994ft}.  As a consequence, it is obvious that the stress tensor does not thermalize to the canonical ensemble.  For example, if we consider a CFT on a spatial circle and prepare an initial state such that $\langle T_{++}(\phi,0)\rangle=f(\phi)$, then at later times we will have $\langle T_{++}(\phi,t)\rangle=f(\phi+t)$. The spatial dependence does not evolve to a constant value compatible with the thermal value.   On the other hand, we know that holographic CFTs at large central charge do thermalize in the sense that in the bulk description highly energetic states collapse to form black holes, and these are dual to thermal states in the CFT.
The question is whether 2d CFTs are compatible with a generalized ETH in which the conserved KdV charges are taken into account.    
  

Our approach is to examine the validity of the standard and  generalized ETH ansatze in the case that $O$ is a primary operator.   The energy eigenstates in a CFT are organized into primaries and descendants.  If the states $|\psi_{a,b}\rangle$ are both primary, then the matrix element is the primary OPE coefficient $C_{\Oc_a O \Oc_b}$, which is part of the CFT data.   These coefficients must respect the ansatz (\ref{eth}), (assuming of course that $|\psi_{a,b}\rangle$ are high energy states) in order for ETH to have a chance of being satisfied.  Specifically, we demand that the primary OPE coefficients obey
\bea\label{ethb}
C_{\Oc_a O \Oc_b} \approx \overline{O} \delta_{ab} + e^{-S\left({E_a+E_b\over 2}\right)/2} f_{O}(E_a,E_b) R_{ab}~,
\eea
where the energy $E$  is related to the scaling dimension $\Delta$ as $E=\Delta-{c\over 12}$.  The energies $E_{a,b}$ must be sufficiently large such that $e^{-S(E_{a,b}}\ll 1$, so that the off-diagonal elements are nonzero but suppressed compared to the diagonal elements.   We are assuming that all operators are normalized such that their 2-point functions on the plane obey $\langle O_a(1)O_b(0)\rangle =\delta_{ab}$.  Our focus in this work is to determine the extent to which this form does or does not hold for high energy descendant states. The above statement of ETH in CFTs in terms of OPE data is the same as in the earlier works, e.g.   \cite{Lashkari:2016vgj}

 Modular invariance and crossing symmetry control the average value of these OPE coefficients in the high energy regime \cite{Kraus:2016nwo,Cardy:2017qhl,Das:2017cnv,Brehm:2018ipf, Romero-Bermudez:2018dim,Hikida:2018khg,Collier:2018exn}, but their fine grained structure is theory dependent.  
 We stress that to address the question of whether some version of  ETH holds in 2d CFT we need to study individual matrix elements; universal CFT results regarding average values are not sufficient.

Assuming that the primary OPE coefficients are compatible with ETH we examine descendant states.  We emphasize that at high energy the vast majority of states are high level descendants \cite{Datta:2019jeo}.  In particular, for states at energy\footnote{The full CFT Hamiltonian is actually $H=L_0+\overline{L}_0$, but   we restrict attention to the holomorphic side.}  $L_0=h$,  the fraction of  states which are primary is
\bea
\frac{N_{\rm prim}}{N_{\rm tot}}\approx  e^{-2\pi (\sqrt{c}-\sqrt{c-1}) \sqrt{h\over 6}}~,
\eea
which is exponentially small as $h\rt \infty$. So a proper test of ETH in 2d CFT requires us to examine matrix elements involving high level descendant states.  These are fixed by the Virasoro algebra in terms of the primary matrix elements, but no systematic attempt has been made to compute them explicitly, as far as we are aware.  The main step forward in this work is that we develop the tools to compute such matrix elements and to study their patterns.    Even apart from ETH considerations, the structure of such matrix elements is a universal fact about 2d CFTs, and so worth exploring for its own sake.

The ETH form (\ref{eth}) is supposed to hold in an orthonormal basis of energy eigenstates.   In 2d CFT, a complication is that the energy spectrum is degenerate by virtue of the Virasoro symmetry.   Descendant states are obtained by acting with raising operators $L_{-n}$ on a primary state $|h\rangle$.  The energy of the descendant state $L_{-1}^{n_1}L_{-2}^{n_2} \ldots .... L_{-j}^{n_j}|h\rangle$ is $L_0 = h+ \sum_j jn_j$, hence there is a degeneracy corresponding to the number of ways of partitioning an integer. Finding an orthonormal basis in terms of this starting basis is a tedious task, essentially intractable beyond the first few levels.  One  needs to compute the matrix of inner products by commuting strings of Virasoro generators past each other and then diagonalize the resulting matrix.   Even if this could be accomplished, the computation of the primary matrix elements in this approach is at least as challenging, as one is required to commute large number of generators past both the primary operator and other generators.    Such a direct approach is therefore hopeless.   

In principle, the optimal basis is one which diagonalizes the KdV charges, since this would offer the most direct way of checking whether generalized ETH holds.   However, finding such a basis is intractable for the reasons noted in the last paragraph, on top of the fact that  explicit expressions for the KdV charges are not even known beyond the first seven.  Instead, we now discuss a way to construct one particular orthonormal basis, and then later turn to the issue of the KdV charges.


We develop an approach that turns out to be well suited to the problem of computing  matrix elements in an orthonoormal basis. The Virasoro algebra for arbitrary central charge $c$ and primary dimension $h$ can be represented in terms of differential operators acting on an infinite collection of oscillator variables $U=\{ u_1 , u_2, \ldots\}$ which live on the complex plane. States are represented as holomorphic functions of the oscillator variables, and there is an inner product under which the representation is unitary (here we assume $h,c >0$).  This representation was employed in \cite{zamolodchikov1986two}, where it was used to find a closed form expression for a four-point Virasoro block in the case that $c=1$ with external dimensions all equal to $h=1/16$.
A similar representation has also been used in the context of matrix models for 2d quantum gravity \cite{dijkgraaf1991loop,Dijkgraaf:2018vnm}.

The oscillator representation has several convenient features.  First, it is very easy to construct an orthonormal basis: one simply takes wavefunctions to be proportional to monomials, $\psi(U) \propto u_1^{n_1} u_2^{n_2}\ldots$.  Second, we can derive  recursion relations for the primary matrix elements that can be solved level-by-level, and which are straightforward to implement numerically.  Thus it becomes tractable to work out large numbers of matrix elements at relatively high level and to examine the patterns that result.

We focus on matrix elements of a primary operator $O$ of scaling dimension $h$ taken between descendant states built on primary operators $\Oc_{U,V}$ of equal dimension $h_U=h_V$.  The latter choice is made so that we can simultaneously study both diagonal and off-diagonal matrix elements.   The results depend on the magnitude of the parameters  $(h_U=h_V,h, c)$; here we summarize a few key results.  First, all of the off-diagonal elements are seen to be nonvanishing;\footnote{Of course, within one degenerate subspace we could always go to a new basis that diagonalizes the matrix; however, this basis would only achieve this goal for one particular choice of primary operator dimension $h$.} we do not see any evidence of selection rules that would  rule out ETH behavior for CFTs.
\begin{figure}[!t]
\begin{center}
	\includegraphics[width=\textwidth]{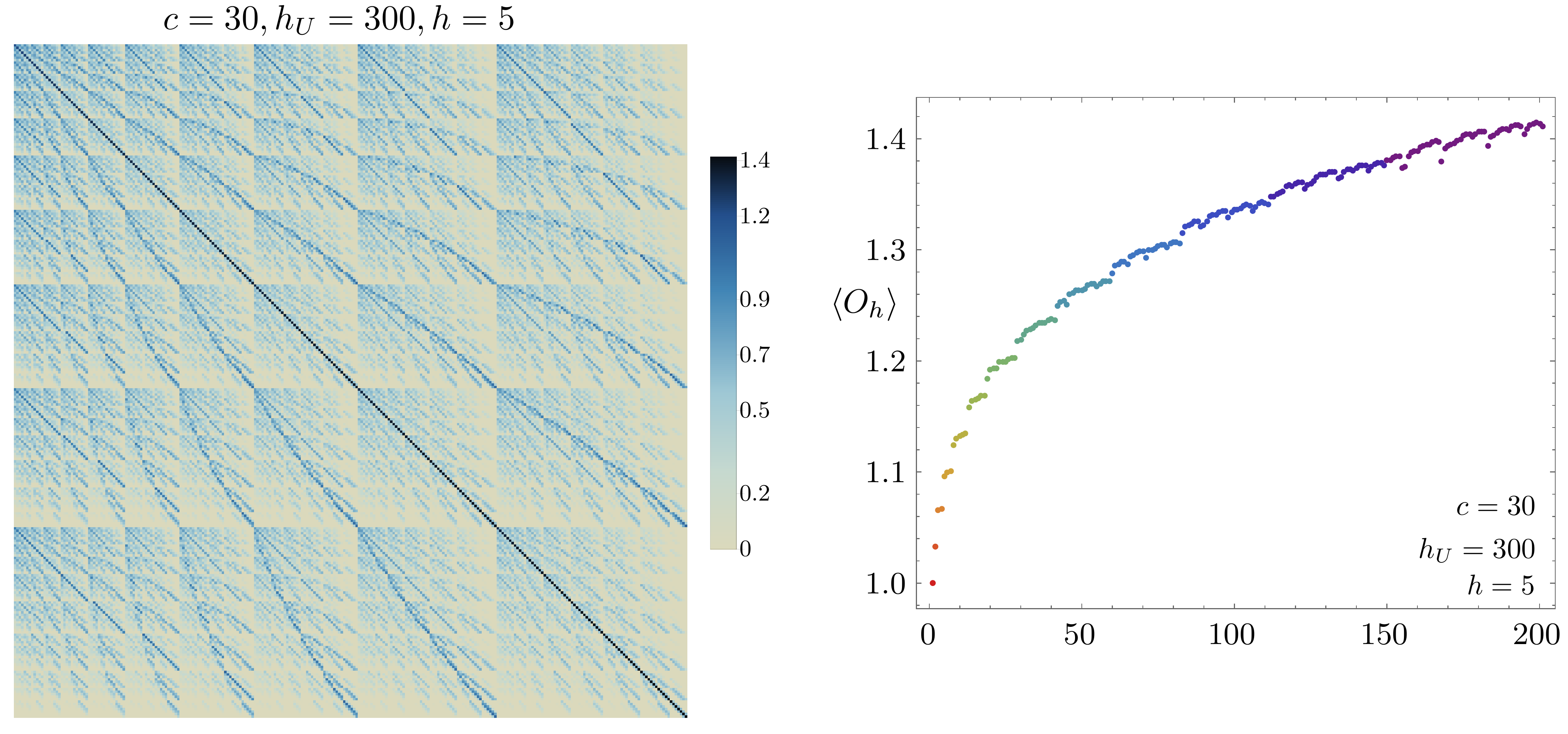}
\end{center}\vspace{-.8cm}
	\caption{[Left] Absolute values of matrix elements, $\vev{h_U, \lbrace m_i \rbrace |O_h | h_V,  \lbrace n_i \rbrace }$, up to descendant level 12  for $h_U=h_V=300$, $h=5$ and central charge $c=30$. [Right] The diagonal matrix elements after filtering out outliers (see text) using $\sum_k\vev{L_{-k}L_k}$ (each descendant level from 1 to 12 is labelled by the colors red to violet).  }
	\label{intro-fig}
\end{figure}

A regime in which  AdS/CFT leads us to expect thermalization to occur is the `heavy-light' regime $h_{U,V},c \gg 1$, $h\sim O(1)$.    As an example we choose $h_U=h_V=300$, $h=5$ and central charge $c=30$.  See Fig.~\ref{intro-fig}.  Suppose we first ignore the existence of the KdV charges, and ask whether ordinary ETH holds.     ETH asserts that the diagonal matrix elements should form a smoothly varying curve, so that at high energy the matrix elements are nearly constant over a window containing many states. Here it is worth emphasizing that in CFT we have degeneracies in the energy spectrum, but that does not mean that we should take ETH to imply exact equality of the corresponding diagonal matrix elements; such a strong condition would imply that essentially all CFTs are incompatible with ETH. Rather, the physical requirement is the one we have just mentioned: two diagonal matrix elements should be nearly equal if the corresponding energies are nearly equal, i.e.~if $\Delta E/E \ll 1$.  In Fig.~\ref{primaryFilter1} (left panel) we see that most of the matrix elements lie along a smooth curve, but there is a significant fraction of outliers.  The presence of these outliers is the signal that ordinary ETH does not hold.  

It is at this point that we need to incorporate the  KdV charges associated to the stress tensor, the simplest of which is essentially $T_2 \sim \sum_k L_{-k}L_k$.  Subsequent charges involve more powers of Virasoro generators and rapidly become more complicated; indeed their explicit form is only known for the 7 lowest lying charges. The results shown in the left panel of Fig.~\ref{intro-fig} are for all states, with no reference to their KdV charges.  However, in generalized ETH we we should only compare matrix elements between states whose conserved charges are nearly equal. 

 We now discuss the implementation of this for $T_2$.  In principle, at each energy level we should choose a basis that diagonalizes $T_2$; however this is  challenging numerically, and so we adopt the simpler but cruder strategy of computing the expectation value of $T_2$ in our existing oscillator basis, laying the foundation for more complete analyses in the future. 
In order to compare states with similar values of $T_2$, we identify outliers as those whose $T_2$  charge lies more than, say, $10\%$ from the minimal value at the given level.   In Fig.~\ref{intro-fig} (right panel) we replot the diagonal primary matrix elements after this cut is made.  The smooth curve that results provides evidence for the conjecture that, at least for this parameter regime, there is no obstruction to generalized ETH once the conserved charges are taken into account.   This claim receives further support from examination of the off-diagonal matrix elements  displayed in Fig.~\ref{intro-fig}, as they are found to be highly suppressed compared to the diagonal matrix elements.   This relative suppression is the key fact needed for operators expectation values to sit at their thermal values ``most of the time". Generic chaotic systems typically exhibit a random matrix structure for the off-diagonal elements.  It is apparent from Fig.~\ref{intro-fig} that some non-random structure is present in 2d CFT, as we identify certain sets of matrix elements that are relatively large.  These special matrix elements can be identified analytically via perturbation theory in $1/h_{U,V}, 1/c$.  This same analysis shows that in the holographic limit the diagonal elements tend to a constant while the off-diagonal elements go to zero, just as we would expect.

Let us denote by ``ETH$_{T_2}$" the statement that matrix elements obey (\ref{eth}) when restricted to states whose $\langle T_2\rangle$ are nearly thermal.   Of course, since there are infinitely many other KdV charges, there are correspondingly further refined versions of ETH.  The general expectation is that generalized ETH will hold to better and better approximation as more charges are included, however the numerical challenge of addressing this is beyond the scope of this work.

We should note that generalized ETH in the sense of (\ref{eth}) will not strictly hold in the sense that the fluctuation in diagonal elements, and the magnitude of off-diagonal elements, are not exponentially suppressed in the entropy.   This is obvious given that the computation of such matrix elements between states in the same module makes no reference to the rest of the theory, and hence cannot know about the total entropy.  The suppression is weaker, as we discuss in the main text.  Of course, for states in different modules we can have an $e^{-S}$ suppression, coming from the overall OPE coefficient.  

We study other parameter regimes as well.  For example, if $c\sim O(1)$, we find that ETH$_{T_2}$ does not hold in the strong sense of applying to all high energy states.  States based on primary operators with $h_{U,V}\sim O(1)$ are found to yield matrix elements in which there is not a universal suppression of the off-diagonal elements relative to the diagonal ones.  On the other hand, typical high energy states in these theories are based on primaries of dimension $h_{U,V} \sim E_{\rm tot}\gg 1$ \cite{Datta:2019jeo}, and here we do get back the ETH structure.   Hence we conclude that ETH$_{T_2}$  holds in the weak sense of applying to most, but not all, of the high energy states.
As we have already noted, the natural next step would be to ask  if the failure of  ETH$_{T_2}$ could be remedied by incorporating more  KdV charges, but we do not address this here.


The remainder of this paper is organized as follows. In Section \ref{sec:osc} we provide a brief review of the oscillator approach which we employ to calculate the matrix elements. The matrix elements are analysed in Section \ref{sec:matrix}. We first study the heavy-light regime which is tractable analytically and then obtain for the matrix elements by numerically solving a linear system. Section \ref{sec:discussion} has our conclusions and a discussion relating our results to the gravity dual. Several details on the oscillator formalism and its utilization to calculate the matrix elements relevant to this work can be found in Appendix \ref{app:osc}.

\section{Virasoro algebra in the oscillator basis}
\label{sec:osc}
Our main task is to compute matrix elements of local operators in an orthonormal basis of energy eigenstates. In two dimensional CFT the Hilbert space is furnished by representations of two copies of the Virasoro algebra
\bea
[L_m,L_n]= (m-n)L_{m+n} +{c\over 12}(m^3-m)\delta_{m,-n}~,
\eea
where $L_n$ are the modes of the stress tensor. The Hamiltonian is $(L_0 + \overline{L}_0)$ \footnote{As noted in the Introduction we supress dependence on the antiholomorphic sector.} and there is a rich structure in a single Virasoro module generated by the action of $L_{n<0}$ on a primary state $|h\>$ with $L_0$ eigenvalue $h.$ For example, stress tensor correlators in a thermal state agree with stress tensor correlators in descendant states with typical features \cite{Datta:2019jeo}.

Our focus will be on unitary $c>1$ CFTs. The usual basis of $L_0$ eigenstates at level $N = \sum_j jn_j$ given by $L_{-1}^{n_1} L_{-2}^{n_2} \ldots L_{-N}^{n_N} |h\>$
is not orthogonal. As detailed in \cite{zamolodchikov1986two} and Appendix \ref{app:osc}, a convenient orthogonal basis is obtained by representing states in a module by wavefunctions of an infinite number of variables $\{u_i,~i=1,2,\ldots\}$ collectively denoted $U.$ A state $|f\>$ in the Virasoro module of the primary $|h\>$ is represented by the wavefunction $f(U) \equiv \<U | f \>.$ At level $N$ descendant states are given by partitions $\{n_j\}$ of the integer $N.$ Corresponding wavefunctions are sums of monomials of the form
\bea
\label{mo}
\<U | h, N, \{n_j\} \> \equiv u_1^{n_1} u_2^{n_2} \ldots u_N^{n_N},~~~~N = \sum_j jn_j~,
\eea
The action of the Virasoro algebra on this basis is given by $\<U|L_k|f\> = l_k f(U)$ where
\begin{equation}
	\label{viro}
	\begin{aligned}
	l_0 & = h+ \sum_{n=1}^\infty n u_n {\p\over\p u_n}, \cr
	l_k &= \sum_{n=1}^\infty n u_n {\p\over\p u_{n+k}} -{1\over 4} \sum_{n=1}^{k-1} {\p^2 \over \p u_n \p u_{k-n} }+(\mu k+i\lambda){\p \over \p u_k}~,\quad\quad k>0 \cr
	l_{-k} &= \sum_{n=1}^\infty (n+k) u_{n+k} {\p\over\p u_{n}} - \sum_{n=1}^{k-1}n(k-n) u_n u_{k-n}+2k(\mu k-i\lambda)u_k~,\quad\quad k>0~.
	\end{aligned}
\end{equation}
Here the central charge and the dimension of the primary are
\bea
c= 1+24 \mu^2,~~h=\lambda^2+\mu^2~.
\eea
In the main text we will take $\lambda$ to be real so that $h\geq \mu^2$.  However, by analytic continuation it is possible to access the region $0 \leq h < \mu^2$ as well, as discussed in Appendix \ref{app:h}.
The state $| h, N, \{n_j\} \>$ has $L_0$ eigenvalue
\bea
l_0 \<U | h, N, \{n_j\} \> = \left( h+ n_1 + 2n_2 + \ldots + Nn_N \right) \<U | h, N, \{n_j\} \>~,
\eea
and the primary state is a constant which we normalize as $\<U|h\> =1.$ For reasons discussed in  Appendix \ref{app:osc} the variables $u_i$ are called ``oscillators" and each is a holomorphic coordinate on a copy of the complex plane. A unitary representation on wavefunctions $f(U),~ g(U) $ realizing Hermitian conjugation relations $l_n^\dagger = l_{-n}$ is given by\footnote{A derivation of the inner product is presented in the Appendix \ref{app:osc}.}
\bea
\label{ip}
\big(f(U),~g(U)\big) =   \int\! [dU] \overline{f(U)}g(U)~,
\eea
with the measure
\bea
[dU] =  \prod_{n=1}^\infty d^2u_n {2n \over \pi} e^{-2n  u_n\ub_n}~.
\eea
Writing each oscillator in polar coordinates, $u_n = r_n e^{i\theta_n}$, it is seen the monomial (\ref{mo}) has non-vanishing inner product only with itself
\bea
\left(u_1^{n_1} \ldots u_N^{n_N},~ u_1^{m_1} \ldots u_N^{m_N} \right) = \d_{n_1,m_1} \ldots \d_{n_N,m_N} S_{1,n_1} \ldots S_{N,n_N}~,
\eea
where
\begin{align}\label{Snorm}
S_{j,k}={2j\over \pi}\int_{\mathbb{C}} du_jd\ub_j ~ e^{-2ju_j\ub_j}|u_j|^{2k}=  (2j)^{-k}\Gamma(k+1)~.
\end{align}
The normalization is such that $(1,1)=1.$

This is the key feature of this formalism since it provides an orthogonal basis of states in a Virasoro module. More specifically, we can write a generic descendant state as a sum of monomials
\bea
\<U|L_{-1}^{n_1} L_{-2}^{n_2} \ldots L_{-N}^{n_N} |h\> = \sum_{\{k_j\}} \cK^{n_1n_2 \ldots n_N}_{k_1k_2 \ldots k_N}~ u_1^{k_1} u_2^{k_2} \ldots u_N^{k_N}~.
\eea
The left hand side is simply computed as $l_{-1}^{n_1} l_{-2}^{n_2} \ldots l_{-N}^{n_N} \cdot 1$ from which the coefficients $\cK$ can be read off using the orthogonality relations.

An immediate application is to the computation of matrix elements of Virasoro generators
\bea
\<h| L_{N}^{m_N} \ldots L_{1}^{m_1} L_{-1}^{k_1} \ldots L_{-N}^{k_N}|h\> = \left( l_{-1}^{m_1} \ldots l_{-N}^{m_N} \cdot 1,~ l_{-1}^{k_1} \ldots l_{-N}^{k_N} \cdot 1 \right)~.
\eea
This is easily computed using the inner product (\ref{ip}) without ever needing to use the commutation relations\footnote{As an aside, this method provides a computationally fast way of obtaining Kac matrices.}. In some cases analytic results as a function of the partitions $\{m_j\},~\{k_j\}$ can be obtained; see Appendix \ref{app:osc} for diagonal matrix elements of $L_n L_{-n}.$

Another simple computation is the two point function. The wavefunction corresponding to a primary displaced from the origin is (please see Appendix \ref{app:wavefunctions})
\bea \psi_h(z,U)=\langle U|O_h(z)|0\rangle = \exp \Big\{{2(\mu-i\lambda) \sum_{n=1}^\infty z^n u_n}\Big\}~.
\eea
With this the two point function is given as
\bea
\< O_h(z_1) O_h(z_2)\> = \int\! [dU] ~  \chi_h(z_1,\Ub) \psi_h(z_2,U) = (z_1-z_2)^{-2h}~,
\eea
where we inserted a complete set of states and defined the wavefunction for the out state $\chi_h(z,\Ub) = \langle 0|O_h(z)|\Ub\rangle,$ which we compute in the   Appendix \ref{app:wavefunctions}. This computation is instructive since the method generalizes to the computation of a generic Virasoro block.


For the discussion of thermalization the main objects we compute in this paper are matrix elements\footnote{We are stripping off the primary OPE coefficient which multiplies the matrix element.}  of primary operators in descendant states $\langle f | O_{h}(z)|g \rangle$ where the corresponding wavefunctions $f(U)$ and $g(V)$ are monomials in representations with dimensions $h_U$ and $h_V.$ The basic object to consider is the matrix element taken between oscillator eigenstates,

\bea
\Omega(z,U,\Vb) =\langle U|O_h(z)|\Vb\rangle~.
\eea
Given this function the general matrix element is extracted via
\bea
\langle f | O_{h}(z)|g \rangle = \int\![dU][dV]\Omega(z,U,\Vb)\overline{f(U)}g(V)~.
\eea
The function $ \Omega(z,U,\Vb)$ can be computed as a solution of the equations
\bea\label{omega}
 \big(\Lc_n^{(z)}+ l_n^{(U)}-\lb_{-n}^{(\Vb)} \big) \Omega(z,U,\Vb) =0~,\quad  n=0, \pm 1, \pm 2~,
\eea
where $ \Lc_n^{(z)} = -z^{n+1}\p_z -(n+1)hz^n$ are the position space conformal generators.
We convert these equations into recursion relations to compute $\Omega(z,U,\Vb)$ in the descendant level expansion. There is no known analytical solution to these recursion relations at all levels.   We therefore solve them numerically for various values of the central charge and conformal dimensions up to descendant level 12 for each state $f(U)$ and $g(V)$.   As we shall see in the next section, it is tractable to obtain analytical solutions in perturbation theory when  $h_{U,V} \gg 1$, possibly along with $c\gg 1$.  This will be useful for interpreting the numerics and obtaining a global picture of the solutions.

\section{Matrix elements}
\label{sec:matrix}
We have seen that the system of equations \eqref{omega} needs to be solved in order to obtain the matrix elements.
In this section, we shall explore the solutions in various regimes of $h_{U,V},h$ and $c$. An analytically tractable case of $h_{U,V}\to \infty$ and $c\to\infty$ is studied first. We then investigate other regimes of parameters by numerically solving \eqref{omega} for the matrix elements.

The $z$ dependence of the function $\Omega$ in \eqref{omega}, can be fixed as
\bea
\Omega(z,U,\Vb) = z^{h_U-h-h_V}F\left(z^{1}u_1, z^{2}u_2, \ldots; z^{-1} \vb_1 ,z^{-2} \vb_2,\ldots \right)~.
\eea
This leads to the system of equations (please see Appendix \ref{app:rec} for details)
\bea\label{sys-eq-2}
\big(-nh -l_0^{(U)}+\lb_0^{(\Vb)} +l_n^{(U)}-\lb_{-n}^{(\Vb)}\big) F(U,\Vb)  =0~,\quad n =\pm 1, \pm 2,\ldots~.
\eea
We shall be interested in the case where the in and out states are based on the same primary, i.e.~$h_U=h_V$, since this lets us access both diagonal and off-diagonal matrix elements.
The function $F(U,\Vb)$  can be written as the following infinite sum
\bea
F(U,\Vb) = \sum_{p,q=0}^\infty F_{p,q}(U,\Vb)~,
\eea
where $F_{p,q}(U,\Vb)$ has level $(p,q)$, defined as  $l_0^{(U)} F_{p,q}(U,\Vb)= (h_U+p)F_{p,q}(U,\Vb)$ and $\lb_0^{(\Vb)} F_{p,q}(U,\Vb)$ $ =(h_V+q)F_{p,q}(U,\Vb)$.  $l_n^{(U)}$ lowers $p$ by $n$ units, and $\lb_n^{(\Vb)}$ lowers $q$ by $n$ units.   Therefore, the equations at fixed level $(p,q)$ are (see Appendix \ref{app:rec} for details)
\bea\label{levpq}
\big[-nh -l_0^{(U)}+\lb_0^{(\Vb)} \big] F_{p,q} +l_n^{(U)}F_{p+n,q} -\lb_{-n}^{(\Vb)}F_{p,q-n} =0~.
\eea
This system needs to be solved for all $n\neq 0$ and all $p,q \geq 0$. It is sufficient to consider the solutions to the equations for $n=-2,-1,0,1,2$;
commutation relations guarantee that the other equations with $|n|>2$ will then be automatically satisfied.
Each $F_{p,q}(U,V)$ is a linear combination of monomials $u_j$ and $\bar v_j$
\begin{align}\label{fc}
	F_{p,q}(U,\Vb) = \sum_{\lbrace m_i \rbrace, \lbrace n_i \rbrace } C^{(h_U,h_V,h,c)}_{\lbrace m_i \rbrace, \lbrace n_i \rbrace} \prod_{j} \bar v_j^{n_j}\prod_{k}u_k^{m_k}~.
\end{align}
Here, $\lbrace m_i \rbrace$ and $\lbrace n_i \rbrace$ are the sets all possible integer partitions of $p(=\sum_k km_k)$ and $q(=\sum_j jn_j)$ respectively. The coefficients $C^{(h_U,h_V,h,c)}_{\lbrace m_i \rbrace, \lbrace n_i \rbrace}$ are finally appropriately normalized to yield the matrix elements in  descendant states
\begin{align}\label{matrix-elem}
	\vev{h_U, \lbrace m_i \rbrace |O_h | h_V,  \lbrace n_i \rbrace }=    C^{(h_U,h_V,h,c)}_{\lbrace m_i \rbrace, \lbrace n_i \rbrace}  \left[   \prod_{j}S_{j,m_j} \prod_{k}S_{k,n_k}   \right]^{1\over 2}   ~.
\end{align}
where $S_{n,m_n}$ are given by \eqref{Snorm}. Therefore, $F(U,\Vb)$ is the key object which encodes the information about the matrix elements.

\subsection{Heavy-light perturbation theory}

\begin{figure}[!t]
	\includegraphics[width=\textwidth]{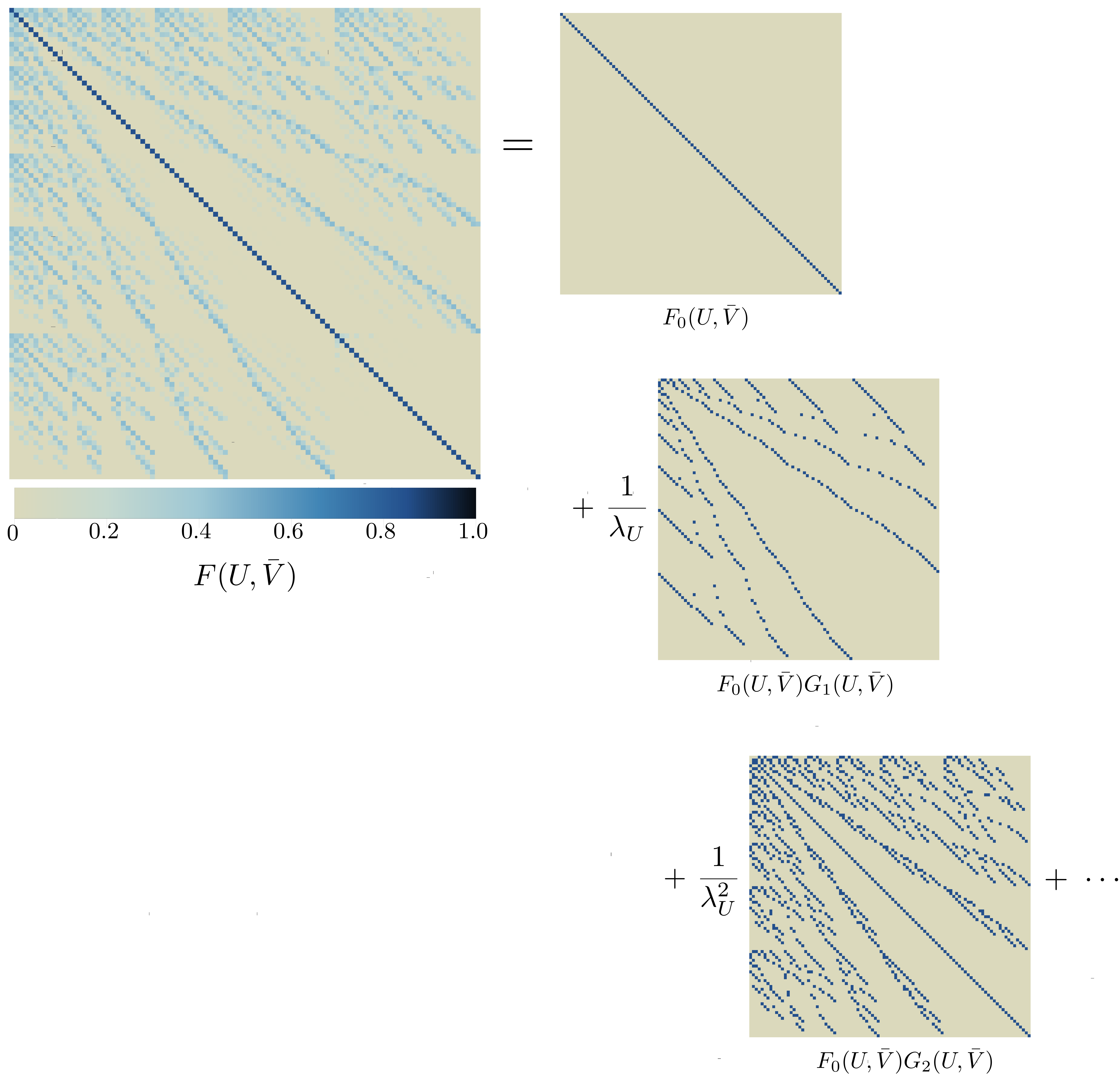}
	\begin{minipage}[c]{0.53\textwidth}\vspace{-6.5cm}
		\caption{
			{\bf Perturbation theory matrices --} The left hand side is the numerical solution of the matrix elements \eqref{matrix-elem} contained within $F(U,\bar V)$ up to descendant level 9 for the parameters $c=5000$, $h_U=h_V=5000$ and $h=1$. The right hand side shows the matrix elements which get activated order-by-order in $1/\lambda_U$ perturbation theory -- equations \eqref{1stord} and \eqref{2ndord}. Clearly these non-zero matrix elements from the first few orders appear more prominently than the rest in the numerical solution.
		} \label{fig:03-03}
	\end{minipage}
\end{figure}

It turns out that the system of equations \eqref{sys-eq-2} for  $F(U,\Vb)$  admits a simple solution in the `heavy-light limit', defined as $\lambda_U=\lambda_V\to \infty ,~ \mu \to \infty$, with the ratio $\alpha=\mu/\lambda_U$ fixed.  We also hold fixed $h$ (the conformal dimension of the probe) and the descendant level. This limit is relevant for AdS/CFT considerations, as discussed in  Section \ref{sec:discussion}.  In this limit,  equation \eqref{sys-eq-2} reads
\begin{align}
	\left(\frac{\partial}{\partial u_n} - 2n \bar{v}_n\right) F_0(U,\Vb) =0~.
\end{align}
This is solved by
\begin{align}\label{0th-ord}
	F_0(U,\Vb) = e^{2\sum_k k u_k \bar{v}_k}	~.
\end{align}
It is worthwhile to note that the above form of $F_0$ is also annihilated by the operator $\big(-l_0^{(U)}+\lb_0^{(\Vb)} +l_n^{(U)}-\lb_{-n}^{(\Vb)}\big)$.  Furthermore, $F_0$ is symmetric under $u_k \leftrightarrow \bar{v}_k$ and it implies that in the leading order only the diagonal matrix elements are non-zero.
This exact solution \eqref{0th-ord} provides motivation for perturbation theory in the parameter $1/\lambda_U$
\begin{align}\label{series-exp}
	F(U,\Vb) = F_0(U,\Vb) \sum_{p=0} \lambda_U^{-p} G_p(U,\Vb)~.
\end{align}
with $G_0=1$.
Plugging this in \eqref{sys-eq-2} and using the explicit form of the Virasoro generators \eqref{viro}, we get the following recursion relation
\begin{align}\label{heavy-rec}
	\frac{\pd G_{p+1}}{\pd u_k} = -\frac{1}{\alpha k+i} \bigg[ & kh   - \sum_{n=1}^\infty  n \left( \bar{v}_n \frac{\pd}{\pd \bar{v}_n} - u_n \frac{\pd}{\pd u_n}\right)    - \sum_{n=1}^\infty  \left( nu_n \frac{\pd}{\pd u_{n+k}} - (n+k)\bar{v}_{n+k} \frac{\pd}{\pd \bar{v}_n}\right)   \nonumber \\
	& + \frac{1}{4} \sum_{n=1}^{k-1} \left(2(k-n)\bar{v}_{k-n}\frac{\pd}{\pd u _n } + \frac{\pd^2}{\pd u_n \pd u_{k-n}} + 2n \bar{v}_n \frac{\pd}{\pd u_{k-n}}\right)   - \mu k \frac{\pd}{\pd u_k} \bigg] G_p~.
\end{align}
The solution at the first order is
\begin{align}\label{1stord}
	G_1(U,\Vb) = h \sum_{k=1}^{\infty} k\left[ \frac{u_k}{\alpha k +i} + \frac{\bar{v}_k}{\alpha k -i}   \right]~.
\end{align}
The higher orders in perturbation theory can be systematically worked out using \eqref{heavy-rec}. However, the expressions get reasonably complicated. At second order, we have
\begin{align}\label{2ndord}
	G_2(U,\Vb) =& \ h \sum_{m,n=0}^\infty mn \left[\frac{u_n}{\alpha n+i} + {\bar{v}_n\over \alpha n -i }\right]\left[\frac{u_m}{\alpha m+i} + {\bar{v}_m\over \alpha m -i }\right]\nonumber \\
	&-  h \sum_{m,n=0}^\infty n^2 \left[\frac{\bar{v}_n q_{m,n}}{\alpha n-i} - {u_n q_{m,n}\over \alpha n +i }\right]\left[\frac{u_m}{\alpha m+i} + {\bar{v}_m\over \alpha m -i }\right]\nn \\
	&-  h \sum_{m,n=0}^\infty m(n+k) \bigg[\left(\frac{u_n q_{m,n}}{\alpha (n+m)+i} - {\bar{v}_{n+m}\over \alpha n -i }\right)\frac{u_m}{\alpha m+i}+ \nonumber\\
	&\qquad\qquad \qquad\qquad\quad  +\left(\frac{\bar{v}_n q_{m,n}}{\alpha (n+m)-i} - {u_{n+m}\over \alpha n +i }\right)\frac{\bar{v}_m}{\alpha m-i} \bigg]\\
	&+ \frac{h}{2} \sum_{m=1}^\infty \sum_{n=1}^{m-1} n (m-n ) \bigg[ \left(\frac{\bar{v}_{k-n}}{\alpha n - i } - \frac{\bar{v}_n}{\alpha(m-n)+i}\right) {u_m \over \alpha m+i} \nonumber\\
	&\qquad\qquad \qquad\qquad\qquad +\left(\frac{u_{m-n}}{\alpha n + i } - \frac{u_n}{\alpha(m-n)-i}\right) {\bar{v}_m \over \alpha m-i} \bigg]\nn .
\end{align}
where $q_{mn}=1-\delta_{m,n}/2$.
It is clear from the above expressions that terms beyond the leading order in perturbation theory activates off-diagonal elements. This can be illustrated using the matrix plots, Fig.~\ref{fig:03-03}.

A similar analysis also holds true when the central charge $c$ and the conformal dimension of the probe $h$ are kept fixed and $h_{U,V}$ are taken to be large. The leading order solution in this case is once again given by \eqref{0th-ord} itself.  Solving for the $F(U,\Vb)$ using the pertubative expansion \eqref{series-exp} again we see that the first order correction is now given by
\begin{align}
	\tilde G_1(U,\Vb) =-i h \sum_{k=1}^{\infty} k\left(   {u_k}  - {\bar{v}_k}    \right)~.
\end{align}
The higher orders can be obtained analogously.

More generally, perturbation theory can also be performed for the case $h_U\neq h_V$ or $\lambda_U\neq \lambda_V$ in the limit $\lambda_{U,V}\to \infty$ and $\mu\to\infty$. The leading order PDE for $F_0(U,\Vb)$ in this case is
\begin{align}
	\left((\mu n + i\lambda_U)\frac{\partial}{\partial u_n} - 2n(\mu n + i\lambda_V) \bar{v}_n\right) F_0(U,\Vb) =0~.
\end{align}
This solution is once again given by a diagonal matrix
\begin{align}\label{off-diag-desc}
	F_0(U,\Vb) = e^{2\sum_k k  \left(\frac{\mu k +i \lambda_V}{\mu k +i \lambda_U}\right)u_k\bar v_k  }.
\end{align}
The perturbative analysis can be carried out now in the two small parameters $1/\lambda_U$ and $1/\lambda_V$, with the ratios $\alpha_{U,V}=\mu/\lambda_{U,V}$ held fixed.

\subsection{Numerics}
\begin{figure}[!b]
	{\centering
		\includegraphics[width=\textwidth]{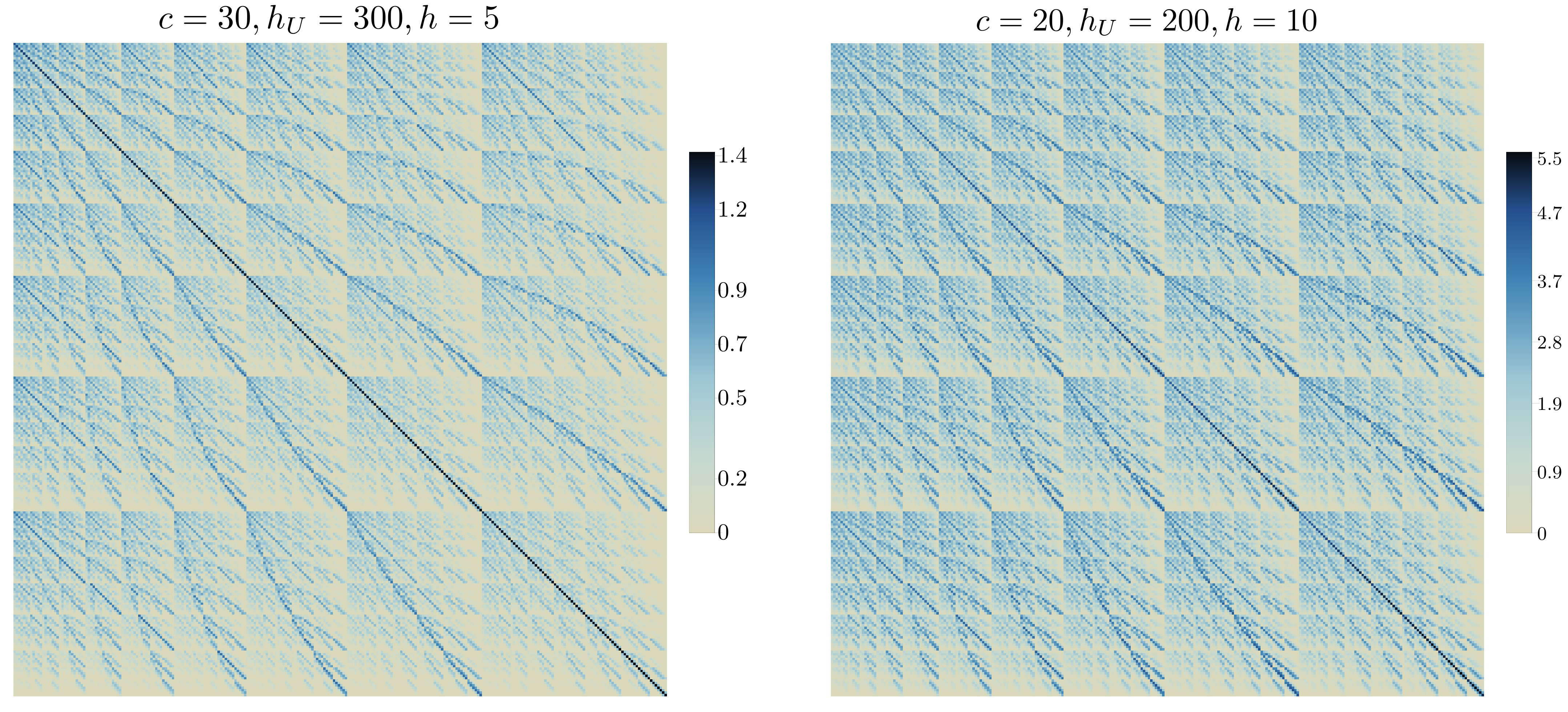} \vspace{-1cm}
	}
	\caption{Absolute values of matrix elements till descendant level 12 at intermediate values of central charge.}
	\label{intermediate}
\end{figure}
\begin{figure}[!b]
	{\centering
		\includegraphics[width=\textwidth]{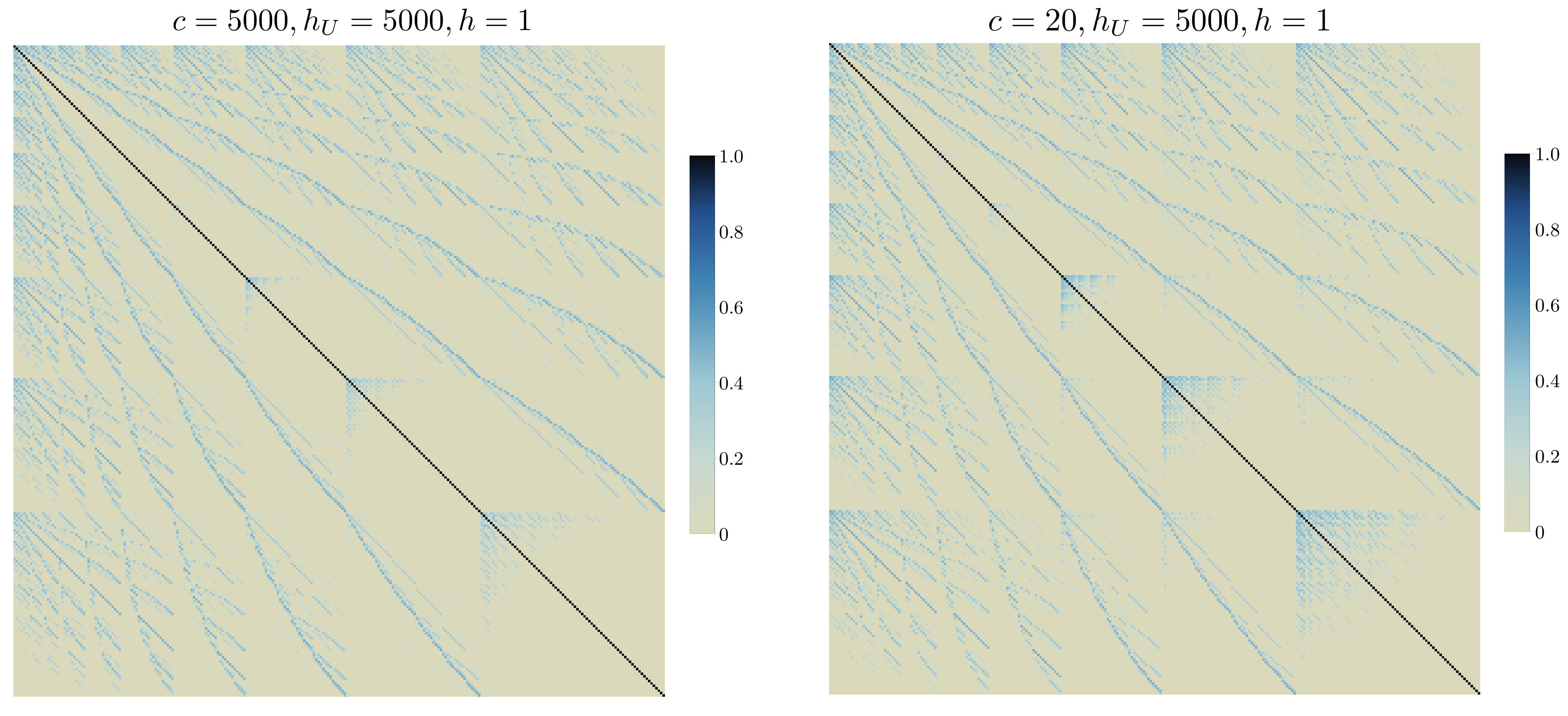} \vspace{-1cm}
	}
	\caption{Absolute values of matrix elements till descendant level 12  for light probes in heavy states.}
	\label{heavylight}
\end{figure}
We now turn to solving the system of equations \eqref{levpq} numerically.
Upon using the differential operator form of the Virasoro generators \eqref{viro} in \eqref{levpq} and plugging in \eqref{fc}, we are led to a sparse linear system of equations for the coefficients $C^{(h_U,h_V,h,c)}_{\lbrace m_i \rbrace, \lbrace n_i \rbrace}$. We start from $F_{0,0}=1$, and then compute the higher level coefficients recursively, up to level $p+q=24$ for some choices of parameters $h_U=h_V,h$ and $c$ \footnote{The matrices in the figures are of dimension $\sum_{j=1}^{12} P(j)=272$ and therefore have 73984 elements each; $P(j)$ are the number of partitions of the integer $j$.}. The system of equations is overconstrained but  consistent solutions can be found.\footnote{We have checked this explicitly by solving some cases exactly upto descendant level 8.} We implement the method  of least squares to solve this linear system numerically. This method solves $A\cdot x=B$ by minimizing the norm $||A\cdot x - B||$.
\begin{figure}[!t]
	{\centering
		\includegraphics[width=\textwidth]{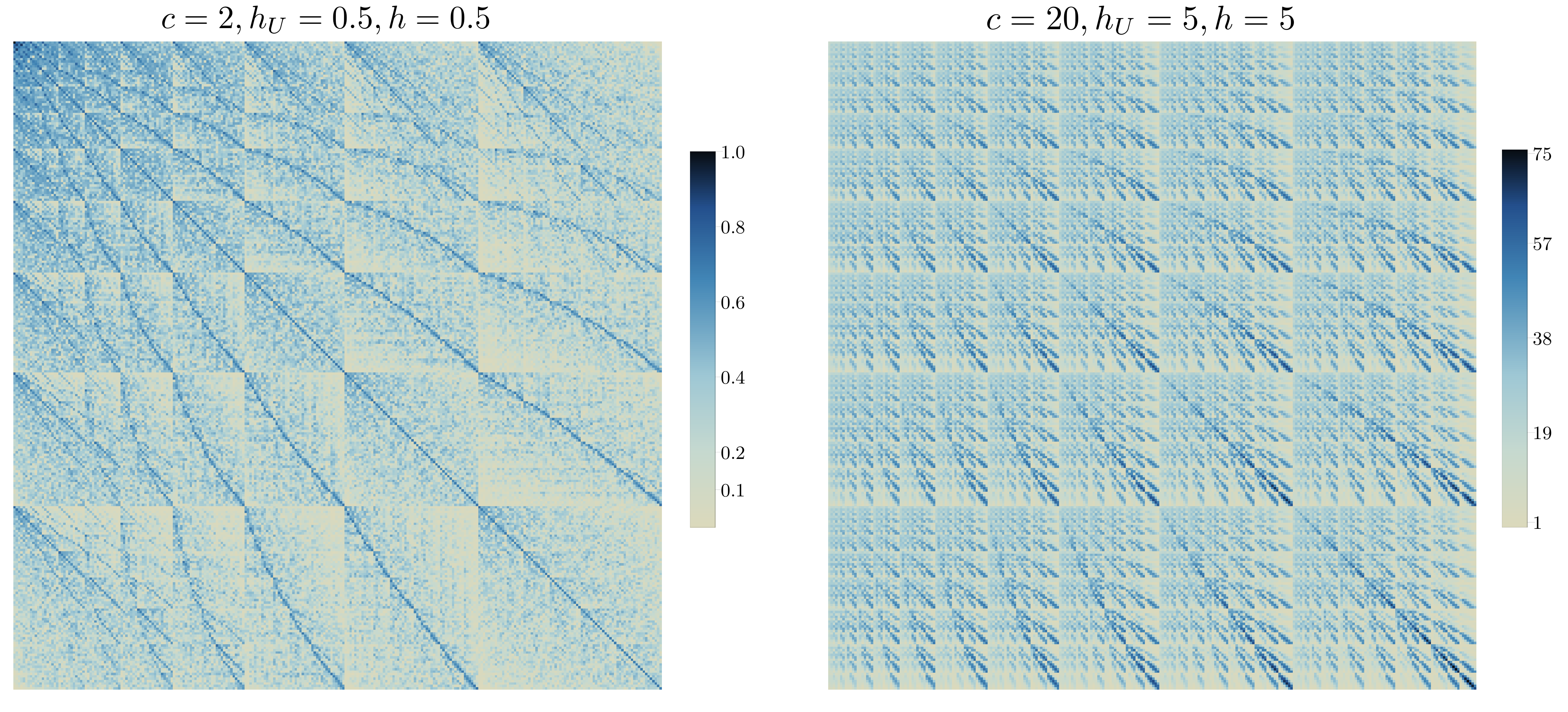}
	} \vspace{-1cm}
	\caption{Absolute values of matrix elements till descendant level 12  for light probes having conformal dimensions of the same order as the primaries in the external states.}
	\label{lightlight}
\end{figure}
\begin{figure}[!t]
	{\centering
		\includegraphics[width=\textwidth]{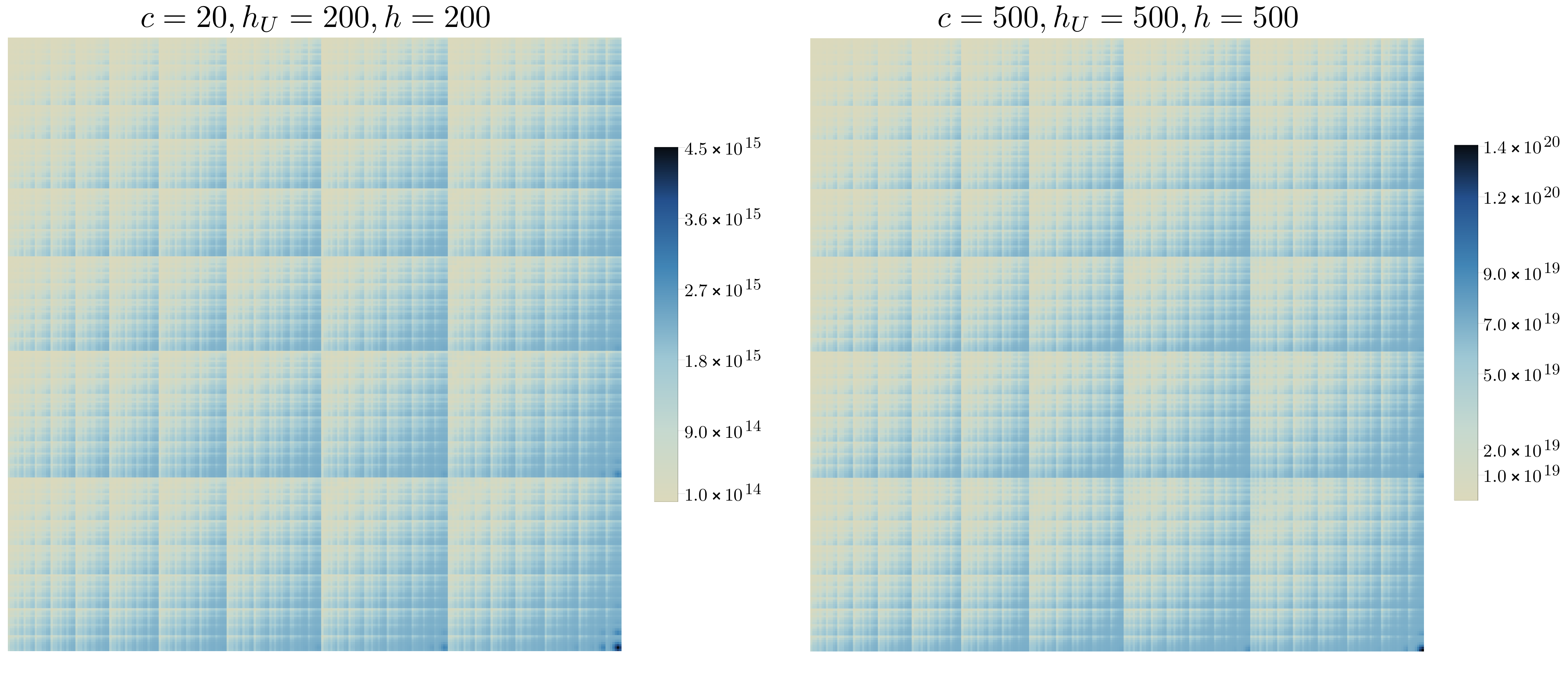}
	} \vspace{-1cm}
	\caption{Absolute values of matrix elements till descendant level 12  for heavy probes in descendants of heavy primaries. The feature of the diagonals being much greater than the off diagonals is clearly lost in this regime.}
	\label{heavyheavy}
\end{figure}
\begin{figure}[!t]
	\begin{center}
		\includegraphics[width=\textwidth]{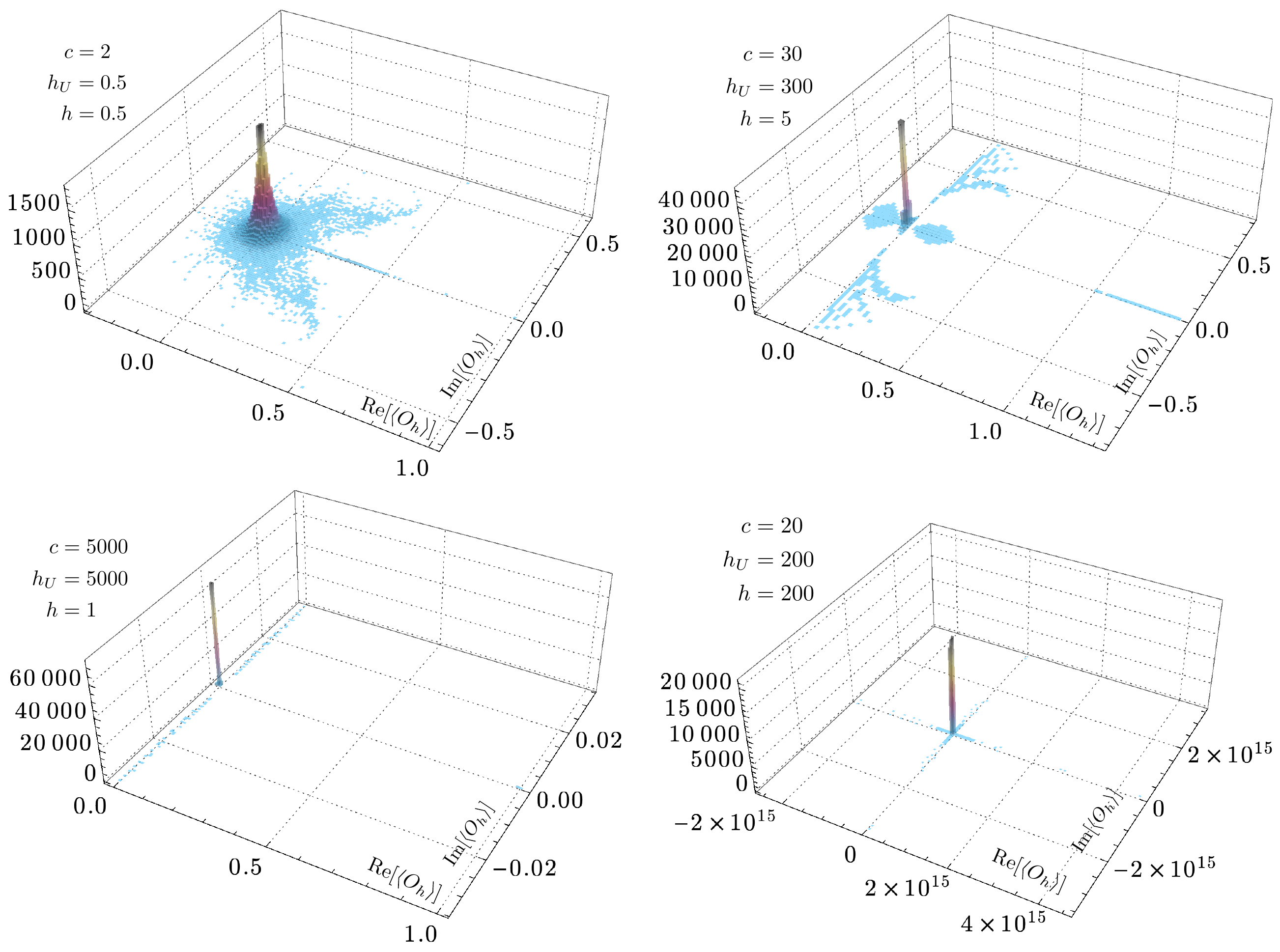}
	\end{center}\vspace{-.7cm}
	\caption{
		{\bf Statistics of matrix elements -- }Histograms of complex values of the matrix elements, till descendant level 12,  for different choices of $h_{U}=h_V$, $h$ and $c$. The off-diagonal elements are peaked about 0. This peak is sharper as the external operator dimensions $h_{U,V}$ are increased.
	} \label{stat}
\end{figure}

It can be seen from the plots of matrices, in Fig.~\ref{intermediate} and Fig.~\ref{heavylight}, that the diagonal elements are of higher magnitude than the off-diagonal elements when $h_{U,V} \gg h$. The mean of the modulus of off-diagonal elements in the cases can be seen to be close to zero and, moreover, their complex values are sharply peaked around zero.
This is also the case where the analytic perturbative solution of the previous subsection applies. Therefore, we conclude that the off-diagonal elements are suppressed \textit{at least by} $1/\lambda_{U,V}$ or $1/\sqrt{h_{U,V}}$ compared to diagonal elements. This situation within a single conformal family in 2d CFTs is in contrast with the general expectation from the ETH ansatz \eqref{eth} which states that \textit{all} off-diagonal elements are exponentially suppressed by the entropy.  On the other hand, recall that we are presently discussing matrix elements between states in the same conformal family.  For states in different families we should remember that the matrix element is multiplied by the corresponding primary OPE coefficient $C_{\Oc_U O_h \Oc_V}$, and these are known, at least on average, to be exponentially suppressed \cite{Brehm:2018ipf,Romero-Bermudez:2018dim,Hikida:2018khg}.

On the other hand, in the regimes where $h_{U,V}$ are of the same order as the conformal dimension of the probe $h$, we do not observe a hierarchy between diagonal and off-diagonal elements --- Fig.~\ref{lightlight} and Fig.~\ref{heavyheavy}. The magnitudes of a considerable number of off-diagonal elements in these cases can be as large as the diagonal elements and their distribution is not sharply peaked either. Furthermore, there also exist diagonal elements with small values.

We therefore conclude that the matrix elements meet the criteria for thermalization in the ``heavy-light regime", i.e.~the limit of light probes and heavy external states. This is expected on general physical grounds and also from holography.  On the other hand, a heavy primary probe holographically corresponds to a very massive scalar field in the bulk, generating significant backreaction, and a much more complicated thermalization process.  It is not surprising that the  ETH ansatz does not apply in this regime.
Finally, the matrices in all cases clearly show a universal repetitive pattern as we move across the sectors indexed by the descendant levels $(p,q)$. This is not the random matrix behavior  appearing in the ETH ansatz.  However, this random matrix structure is not the key  requirement  for thermalization: what matters most is that the off-diagonal elements are non-zero but numerically suppressed compared to the diagonal elements --- the histograms in Fig.~\ref{stat} demonstrate these statistics. 
{For an initial state being a superposition of states from a single Verma module, the power law suppression of the off-diagonal elements (instead of exponential) will lead to comparatively larger fluctuations than what is expected from the ETH ansatz. }

%

\subsection{Restriction to states with fixed KdV charge $T_2$ }

As discussed in the introduction, due to the existence of conserved KdV charges, we need to sharpen what we mean by thermalization in 2d CFT.
We can refine our ETH ansatz by only including states whose KdV charges are nearly thermal (as defined by the Boltzmann ensemble).  We restrict attention to the lowest KdV charge $T_2 = \sum_{k=1}^\infty L_{-k}L_k$, and use that to define a refined ETH$_{T_2}$ ansatz.   If $|\psi\rangle$ is a level-$j$ descendant then $L_{k>j}|\psi\rangle=0$, hence it is sufficient to consider  $\vev{T_2} \equiv \sum_{k=1}^j \langle \psi|L_{-k}L_k|\psi\rangle$.  We discard states that are outliers with respect to this quantity.

\begin{figure}[!t]
	{\centering
		\includegraphics[width=\textwidth]{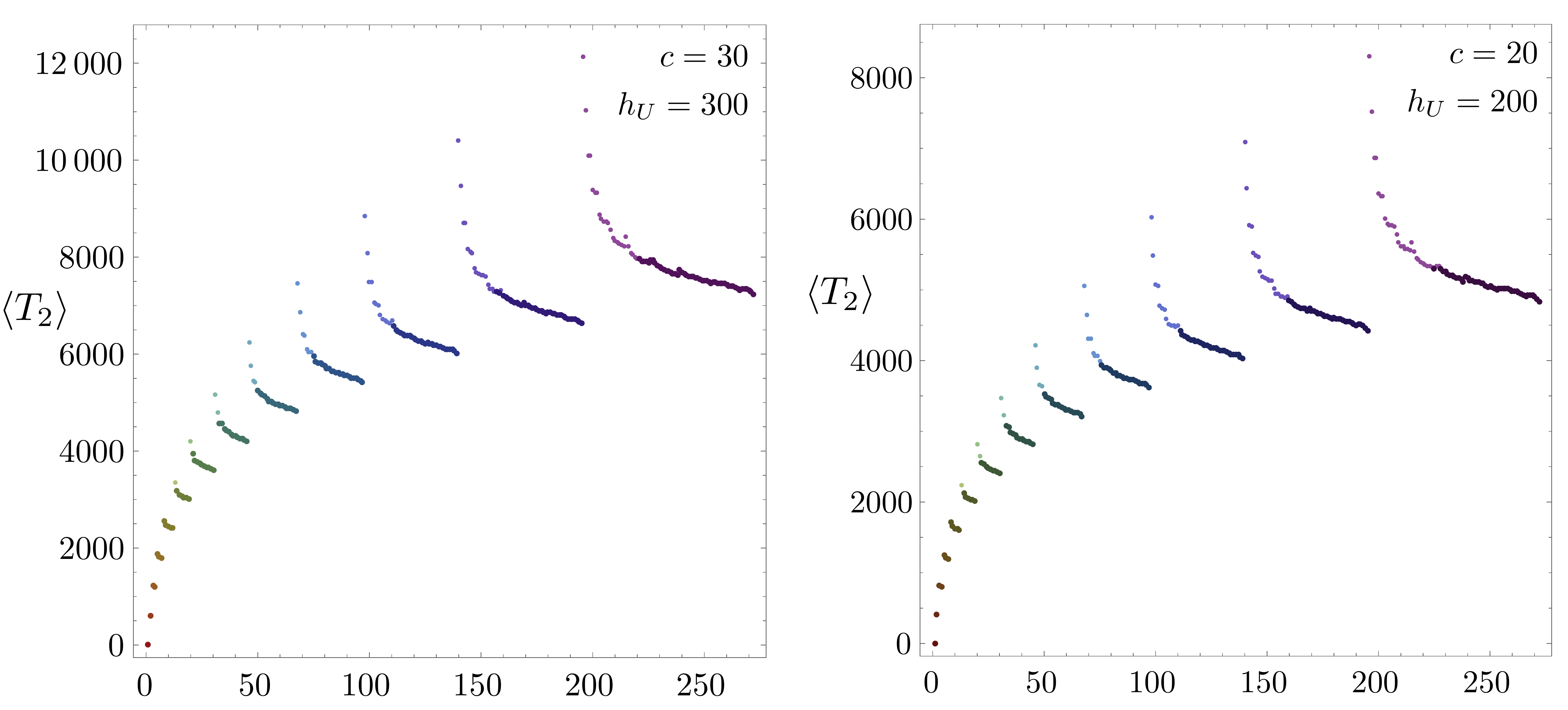}
	}\vspace{-.8cm}
	\caption{Diagonal elements of the operator $T_2=\sum_{k>0} L_{-k}L_k$ from descendant levels 1 to 12 (marked by colors red to violet). The points marked by darker colors are close to the typical value while the ones in a lighter color are outliers.}
	\label{kdvFilter}
\end{figure}
\begin{figure}[!t]
	{\centering
		\includegraphics[width=\textwidth]{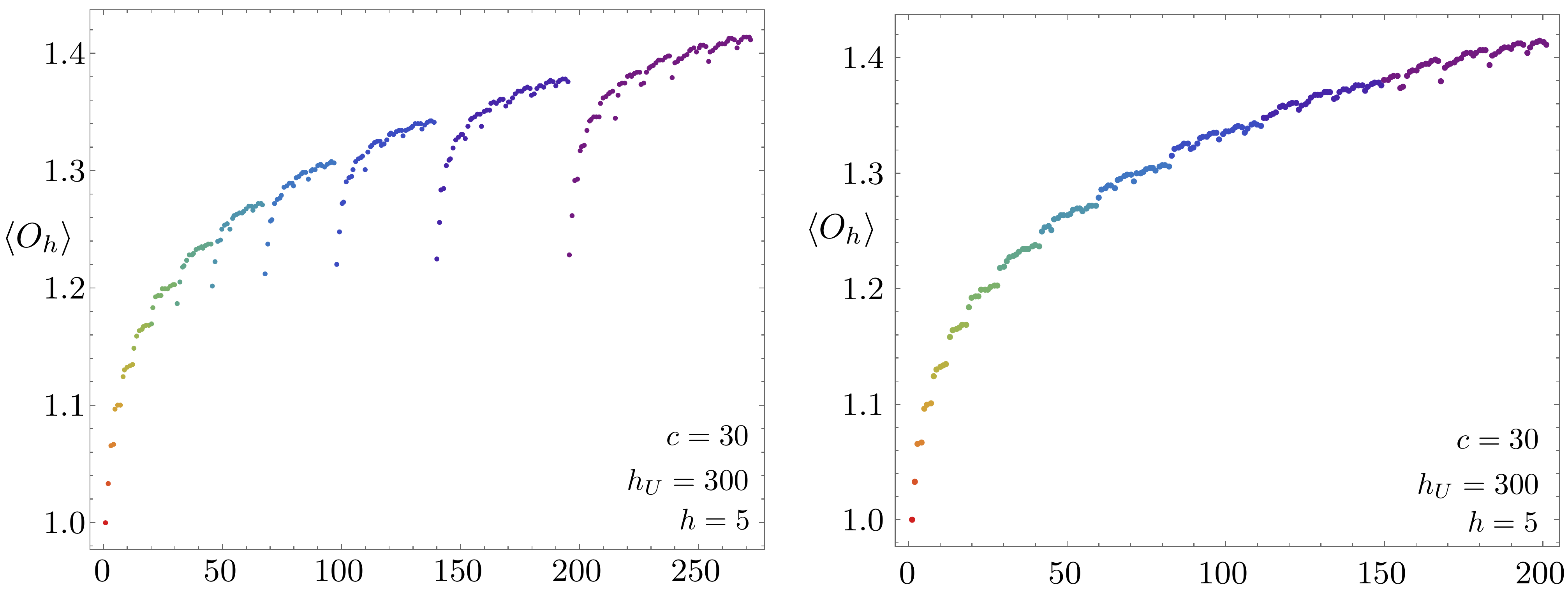}
	}\vspace{-.8cm}
	\caption{[Left] Diagonal elements of the primary $O_h$ from descendant levels 1 to 12 (marked by colors red to violet) for $c=30, h_U=h_V=300$ and $h=5$. [Right] The subset of diagonal elements after filtering out   outliers using the expectation value of $T_2$. }
	\label{primaryFilter1}
\end{figure}
\begin{figure}[!t]
	{\centering
		\includegraphics[width=\textwidth]{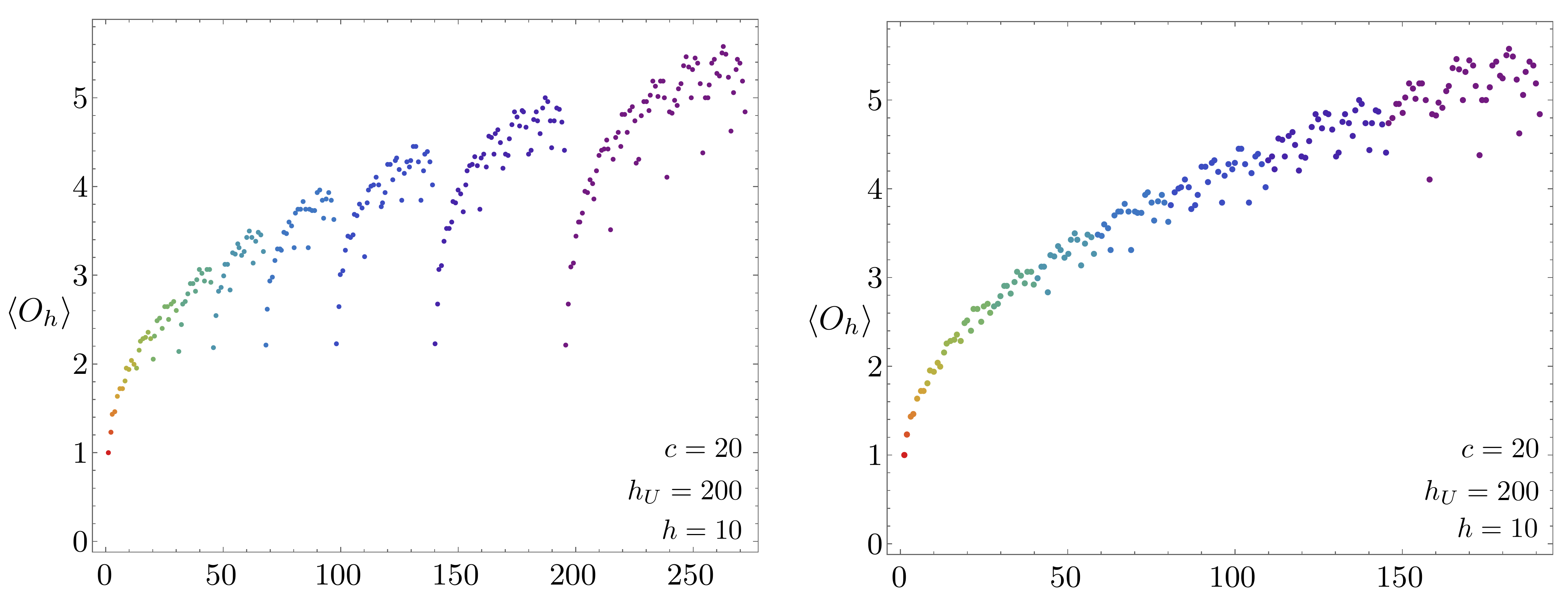}
	}\vspace{-.8cm}
	\caption{[Left] Diagonal elements of the primary $O_h$ from descendant levels 1 to 12 (marked by colors red to violet) for $c=20, h_U=h_V=200$ and $h=5$. [Right] The subset of diagonal elements after filtering out   outliers using the expectation value of $T_2$. }
	\label{primaryFilter2}
\end{figure}

$T_2$ is easy to compute using the expectation values of $L_{-k}L_k$ in an arbitrary descendant state; these are evaluated in Appendix \ref{app:lklk}.  The final results are \eqref{osc-ev} and \eqref{osc-ev-even} and we quote them here. In a descendant state, $\Psi = u_1^{m_1}u_2^{m_2}u_3^{m_3}\cdots$, of a primary of conformal dimension $h_U$, we have the following expectation value
\begin{align}\label{osc-ev-s}
	\vev{l^{(U)}_{-k}l^{(U)}_k}_\Psi \stackrel{{\rm odd}\,k}{=} &  \sum_{p=1}^\infty p(p+k)m_{p+k} {(m_p+1)} +\frac{1}{2} \sum_{p=1}^{k-1}p(k-p) m_p m_{k-p}    +(2kh_U+ \tfrac{c-1}{12}(k^3-k))m_k ,\nn \\
	\vev{l^{(U)}_{-k}l^{(U)}_k}_\Psi  \stackrel{{\rm even}\,k}{=} & \sum_{p=1}^\infty p(p+k)m_{p+k}   {(m_p+1)} + {\frac{k^2}{16} m_{k/2}(m_{k/2}-1) }  +\frac{1}{2} \sum_{p=1}^{k/2-1}p(k-p) m_p m_{k-p}  \nn \\ & + \frac{1}{2} \sum_{p=k/2+1}^{k-1}p(k-p) m_p m_{k-p} +(2kh_U+ \tfrac{c-1}{12}(k^3-k))m_k~.
\end{align}
For some specific choices of parameters, we obtain results like that shown in Fig.~\ref{kdvFilter}.  Most of the microstates yield a $\vev{T_2}$ near a typical value, which is the thermal value, but we also observe a significant number of outlier states.     We choose a  rule regarding which outlier states to discard ---  those whose $\vev{T_2}$ value departs by more than 10$\%$ from the smallest value at that level.\footnote{We could also compare $\vev{T_2}$ to the thermal value, and discard states which deviate too much from that.  This is equivalent to our rule at high levels, but is less convenient to implement.}

Having discarded the outlier states, we compute the matrix elements of a primary operator and ask whether they exhibit ETH behavior, i.e. that the diagonal matrix elements lie on a smooth curve and that the off-diagonal elements are nonzero but numerically suppressed compared to the diagonal values.     The results are shown in Fig.~\ref{primaryFilter1} and Fig.~\ref{primaryFilter2}.   The crucial feature of note is that the states which were deemed outliers with respect to the stress tensor are the same states that are outliers with respect to the primary operator.  \ So for this example, we see that although ETH does not hold, the refined version ETH$_{T_2}$ does.

Let us also comment on the size of the fluctuations of the matrix elements around the smooth interpolating curve.  In generic chaotic quantum systems the fluctuations are typically found to be exponentially suppressed in the system size, i.e. entropy \cite{2016AdPhy..65..239D,deutsch2018eigenstate}.    This cannot be the case here, since our computations only refer to a single Virasoro representation and making no reference to the ``entire" physical system.  Rather, one might expect the fluctuations in our case to be  suppressed by the degeneracy of the representation under study.  It is difficult to accurately characterize the size of the fluctuations given that we are limited by numerics to relatively low level.   Regardless, it is clear that the fluctuations are larger than in a generic chaotic system, illustrating again the distinction with 2d CFTs.  


\section{Discussion}
\label{sec:discussion}

ETH is a strong statement about  matrix elements in chaotic quantum theories.  We have undertaken what we believe is the first systematic study of individual matrix elements in 2d CFTs in the high energy regime relevant to a thermalizing system.   This is possible since the vast majority of high energy states are high level Virasoro descendants, and the corresponding matrix elements  can be determined (up to the overall primary OPE coefficient) using the Virasoro algebra.  What is remarkable is that although these matrix elements are determined by symmetry, they still exhibit a rich and complicated structure.

\subsection{Summary of results and implications for ETH}

We now survey our results in the various parameter regimes, and their implications for ETH.  The most straightforward case is the heavy-light limit, defined by $h_U, h_V, \rt \infty$ with $h_{U,V}/c$ and $h$ fixed.   After removing states in which $\langle T_2\rangle$  is non-thermal, we observed  ETH like behavior for the matrix elements: the diagonal elements lie along a smoothly varying curve, and off-diagonal are nonzero but numerically suppressed relative to the diagonal elements. This is consistent with our refined  ETH$_{T_2}$  ansatz. The off-diagonal elements do not exhibit the random matrix structure that has been observed in various chaotic systems. Instead, we observe some structure in the matrix elements that can be understood analytically in a $1/\lambda_{U,V}$ or $1/c$ expansion.  To establish thermalization, what matters is the systematic suppression of the off-diagonal elements; the random matrix structure is not needed.   The heavy-light limit is relevant for AdS/CFT considerations, and is discussed further in this context below.

In the ETH literature it is emphasized that the ETH ansatz for matrix elements $\langle \psi_a |O|\psi_b\rangle$ is only expected to hold if $O$ is a ``few body operator".  For instance, if the system is a collection of interacting particles, $O$ should only involve  a small number of the particles.
In our 2d CFT setting, the ``few body" property appears to be analogous to the requirement  $h_O \ll h_{U,V}$.   Our numerics show that when $h_O \sim h_{U,V}$ there are off-diagonal elements that are unsuppressed, and further the diagonal elements are not slowly varying.  The AdS version of this is that the scalar field dual to $O$ is sufficiently massive that it backreacts significantly on the background geometry.  Clearly, the approach to equilibrium is much more complicated in this case.

Another case to consider is when $c, h_{U,V}, h_O \sim O(1)$, relevant for generic non-holographic CFTs.  In this case, we need to consider very high level descendant states, since ETH is a statement about high energy eigenstates.  Our numerics (see Fig.~\ref{lightlight}) show that ETH/ETH$_{T_2}$ does not hold in this regime, as there is no universal suppression of the off-diagonal elements.  So these CFTs do not obey ETH/ETH$_{T_2}$ in the strong sense of applying to all high energy states.   On the other hand, we emphasize that the states just mentioned are highly atypical.  As discussed in \cite{Datta:2019jeo}, a typical state at energy $h_{\rm tot}$ is  a level $h_{\rm tot}/c$ descendant of a dimension $ {c-1 \over c}h_{\rm tot}$ primary.   For such large dimension primaries we did find a universal suppression of the off-diagonal element using perturbation theory.   We therefore conclude that a weak version of ETH holds in these theories: it is valid in most states, but not all. As discussed in the introduction, a stronger version might hold once additional conserved KdV charges are incorporated, and this would be an interesting --- albeit numerically challenging --- avenue to pursue. 
  
One can contemplate other applications of our methods.   The oscillator formalism used here can be implemented to compute Virasoro conformal blocks, and may be helpful in accessing new parameter regimes. The matrix elements considered here are also relevant for calculating the spectrum of the Hamiltonian deformed by a primary operator. 2d CFTs with extended symmetries, such as supersymmetry or higher spin symmetry, could also be studied given an oscillator representation of the extended algebra.   Another direction would be to study unitary minimal models, where one does not expect ETH to apply.    Taking $\mu$ to be pure imaginary implies $c<1$; however this yields a non-unitary representation of the Virasoro algebra.   Therefore, modifications to the oscillator formalism are required to study unitary theories with $c<1$.

\subsection{Thermalization in AdS/CFT}

At large $c$ the dual bulk AdS$_3$ theory provides a simple description of thermalization, and it is useful to recall  how it emerges from CFT  considerations.   In AdS$_3$ sufficiently energetic collapsing matter will form a BTZ black hole \cite{Bhattacharyya:2009uu} (this was studied from a CFT perspective in \cite{Anous:2016kss}).  The BTZ black hole has a constant boundary stress tensor, which is used to compute the mass and angular momentum of the black hole.  More generally, if the collapsing matter is inhomogeneous in the boundary direction the collapse will result in a boundary stress tensor with nontrivial profiles described by functions $T_{++}(x^+)$ and $T_{--}(x^-)$. See \cite{deBoer:2016bov} for related discussion.  This is the bulk version of the statement that the stress tensor does not fully thermalize in 2d CFT.   In this classical large $c$ limit this is a coordinate dependent statement: under coordinate transformations that act conformally on the boundary the stress tensor components transform with a Schwarzian derivative term, and we can always perform such a transformation so that the new stress tensor components are constant and hence thermal.  In our previous discussion of outlier states we also used the KdV charge $\sum_k L_{-k} L_k$ to characterize states as being thermal or non-thermal.   However, if we take $c\rt \infty$ holding everything else fixed, then all states become thermal according to this criterion \cite{Basu:2017kzo,Dymarsky:2018lhf,Maloney:2018hdg}.  Therefore, a bulk description of non-thermal states with constant stress tensor is only available at the quantum level; in particular one would need to set up atypical quantum states of the boundary gravitons.

We now add in a probe scalar field, dual to a CFT primary operator of dimension $h_\phi\sim O(1)$.   In the bulk, we imagine adding to the black hole background some scalar field profile at $t=0$ with normalizable falloff at the boundary.  Under time evolution the scalar field will decay by ``falling through the horizon", eventually leading us back to the pure gravity solution, in accordance with the no-hair theorem.   The last stage of the decay is exponential in time, with a timescale set by the lowest quasinormal mode \cite{Horowitz:1999jd,David:2015xqa}.  On the CFT side, this will correspond to a one-point function for the dual CFT operator $O_\phi$ that decays to zero at late time.

Now we examine this process on the CFT side in light of our results.  We first need a description of the state corresponding to the black hole plus scalar field profile.   The black hole itself is described by a typical descendant of a heavy primary operator $\Oc_{\rm BH}$ with scaling dimension $h_{\rm BH} \sim c$ \cite{Datta:2019jeo}.   There are of course many such operators since their multiplicity must account for the black hole entropy.  We then think of fusing such an  operator with the scalar field operator $O_\phi$ to obtain operators describing a black hole plus scalar.  There will again be many such operators with closely spaced dimensions; we call them $\Oc_{{\rm BH}\phi,i}$, with the index $i$ labelling distinct primary operators.

The initial black hole plus scalar field state is then
\bea \label{init}
|\psi\rangle = {1\over \sqrt{\mathcal{N}}}\sum_i \sum_{\{ m_a\} }  \psi_{i,\{ m_a\}} | \Oc_{{\rm BH}\phi,i};\{ m_a\} \rangle~.
\eea
The $\{m_a\}$ labels typical descendant states according to our oscillator conventions and  $1/\sqrt{\mathcal N}$ is the appropriate normalization depending on the states present in the superposition\footnote{If each individual eigenstate in the superposition is normalized to unity, the norm ${\cal N}$ of the unnormalized initial state  is simply the number of states that enter the linear combination.}.  The decay process corresponds to studying $\langle \psi|O_{\phi}(t)|\psi\rangle$ for $t>0$.  Using our results, we can easily compute the part of this expectation value which is determined by Virasoro symmetry.  A major simplification is that at large $c$ the only nonzero matrix elements are those between states with the same descendant labels $\{m_a\}$.  Further the diagonal elements are easily extracted from \eqref{off-diag-desc}.    What  remains is an expression in terms of the primary 3-point coefficients, $C_{ij\phi}= \langle  \Oc_{{\rm BH}\phi,i}|O_{\phi} (0)|  \Oc_{{\rm BH}\phi,j} \rangle$.  Their precise values depend on the particular CFT in question, however their average values\footnote{The averaging being taken over all heavy operators within some narrow scaling dimension window.}   are fixed by modular invariance \cite{Brehm:2018ipf,Romero-Bermudez:2018dim,Hikida:2018khg}.   This average value is exponentially suppressed, hence compatible with the ETH ansatz. There are also exponentially many such operators, which is what allows us to create an initial state with nonzero $\langle O_\phi\rangle$.  As long as the variance of the primary OPE coefficient around the average is not too large,  thermalization then proceeds by the same mechanism as in generic chaotic systems.  The expectation value is given by a sum of exponentials with closely spaced frequencies, $\langle O_\phi(t)\rangle \sim \sum_{ij} c_{ij} e^{-i (h_i-h_j)t}$, and then phase decoherence causes a decay, yielding a value that  tends to zero in the large $c$ limit in which the level spacing vanishes.

This discussion confirms what we expect --- that what we know about 2d CFT at large $c$ is compatible with the bulk picture of thermalization via matter falling through black hole horizons.

\section*{Acknowledgements}
We would like to thank Pawel Caputa, Alex Edison, Ben Michel, Greg Moore and Tadashi Takayanagi for discussions.
This work used computational and storage services associated with the Hoffman2 Shared Cluster provided by the UCLA Institute for Digital Research and Education's Research Technology Group. S.D.~thanks the Yukawa Institute for Theoretical Physics at Kyoto University, where a part of this work was completed during the workshop YITP-T-19-03 `Quantum Information and String Theory 2019' and for an opportunity to present this work.  P.K.~is supported in part by NSF grant PHY-1313986.
	
\appendix	
\section{Oscillator formalism}
\label{app:osc}
In this work we have found it computationally convenient to employ a representation of the Virasoro algebra in terms of differential operators acting on functions that depend holomorphically on certain ``oscillator variables" \cite{zamolodchikov1986two}.  In this appendix we provide a systematic discussion of these methods.  Before turning to the Virasoro case we will warm up with SL(2,$\R$), which is simpler since only a single oscillator variable is needed.

We consider the Virasoro algebra
\bea
[L_m,L_n]= (m-n)L_{m+n} +{c\over 12}(m^3-m)\delta_{m,-n}~,
\eea
or its SL(2,$\R$) subalgebra generated by $\{L_{-1},L_0,L_1\}$.   A Virasoro primary operator of scaling dimension $h$ is written as $O_h(z)$ where $z$ is a coordinate on the plane.\footnote{All dependence on anti-holomorphic quantities will be suppressed throughout.}    It obeys
\bea
[L_n,O_h(z)]= -\Lc_n O_h(z)~,\quad \Lc_n = -z^{n+1}\p_z -(n+1)hz^n~.
\eea
The generators $\Lc_n$ obey the Virasoro algebra with $c=0$,
\bea
[\Lc_m,\Lc_n]=(m-n)\Lc_{m+n}~.
\eea

\subsection{SL(2,$\R$) }

We wish to represent the SL(2,$\R)$ algebra in terms of differential generators acting on holomorphic functions $\psi(u)$ of the oscillator variable $u$ \cite{Bargmann:1946me}. We take $ l_n = u^{1-n} \p_u + (1-n)h u^{-n}$, i.e.
\bea
 l_1 =\p_u~,\quad   l_0 = u\p_u +h~,\quad l_{-1}= u^2\p_u +2hu~,
 \eea
which obey $[l_m,l_n]=(m-n)l_{m+n}$. To build unitary representations we need an inner product that implies the adjoint relations $l_n^\dagger =l_{-n}$.  This is provided by an integral over the unit disk $D=\{ u \in \C, |u|<1\}$ as
\bea
 (f,g) = \int_D [d^2u] \overline{f(u)}g(u)~,\quad  [d^2u] = {2h-1 \over 2\pi} {d^2u \over (1-u\ub)^{2-2h}}~.
\eea
Convergence requires $h>1/2$.  Our convention is that $d^2u = 2dx dy$ with $u=x+iy$, so that $(1,1) = \int_D [d^2u]~ 1  = 1 $.   More generally
\bea\label{uint}
(u^m,u^n) = {n! \over (2h)_n}\delta_{m,n}~.
\eea
Here $(a)_n = \Gamma(a+n)/\Gamma(a)$ is the Pochhammer symbol. Another useful result is
\bea
\big( (l_{-1})^m\cdot 1,  (l_{-1})^n \cdot 1\big) = (2h)_n n! \delta_{m,n}~.
\eea

The basic idea is to represent states in terms of wavefunctions.   The primary state $|h\rangle = O_h(0)|0\rangle$ obeys  $L_0|h\rangle = h |h\rangle$, $L_1 |h\rangle =0$ and so is represented by the unit function $\psi_h(u)=1$.     Other states are obtained by acting with products of $L_{-1}$ on the primary states, and their corresponding wavefunctions are given by the correspondence $L_{-1} \leftrightarrow l_{-1}$.  Wavefunctions are understood as  inner products, $\psi(u)=\langle u|\psi\rangle$, where $|u\rangle$ corresponds to a $u$ eigenstate. Accordingly, $\langle u|L_n|\psi\rangle = l_n \psi(u)$, etc.    We write $\overline{\psi(u)} = \overline{\langle u|\psi\rangle}= \langle \psi|\ub\rangle$, and define the overline to act on the oscillator variable as $u\rt \ub$ but on   the real space coordinate $z$ as an inversion, $\overline{O_h(z)} = (\p_z z')^h O_h(z') $ with $z'=1/z$. The latter follows from the fact that on the plane in and out states are interchanged by coordinate inversion.  We also note that the barred generators  $\lb_n$, which act on anti-holomorphic functions of the oscillator variable,  are given by
$ \lb_n = \ub^{1-n} \p_{\ub} + (1-n)h \ub^{-n}$. We then have
\bea \langle \psi|L_n |\ub\rangle = \overline{\langle u|L_{-n}|\psi\rangle} = \overline{ l_{-n} \psi(u)}=\overline{l}_{-n}\psi(\ub)~.
\eea

The wavefunction corresponding to a primary operator displaced from the origin is
\bea  \psi_h(z,u) = \langle u| O_h(z)|0\rangle~.
\eea
To determine its form we use $L_n|0\rangle =0$ to write
\bea\label{psieq}   0&=&\langle u| L_n O_h(z) - [L_n,O_h(z)]|0\rangle \cr
& =& (l_n + \Lc_n)\psi_h(z,u)~.
\eea
These represent three differential equations which are easily solved to yield (up to normalization)
\bea
\psi_h(z,u) =(1-zu)^{-2h}~.
\eea
Similarly,
\bea \chi_h(z,\ub) = \langle 0|O_h(z)|\ub\rangle
\eea
obeys
\bea (- \Lc_n + \lb_{-n} ) \chi_h(z,\ub) =0~,
\eea
which are solved as
\bea \chi_h(z,\ub) = (z-\ub)^{-2h}~.
\eea
Alternatively, it follows from our definitions that $\chi_h(z,\ub) =z^{-2h}\psi_h(z^{-1} ,\ub)$.

Acting with two primary operators gives us the wavefunctions
\bea \psi_{h_3}(z_1,z_2,u_3) &=& \langle u_3 |O_{h_1}(z_1)O_{h_2}(z_2)|0\rangle~, \cr
 \chi_{h_3}(z_1,z_2,\ub_3) &=& \langle 0 |O_{h_1}(z_1)O_{h_2}(z_2)|\ub_3\rangle~.
\eea
These represent the $h_3$ state that appears in the tensor product of the corresponding $h_1$ and $h_2$ states.   By the same logic as above, these obey
\bea && \big(l_n^{(3)}+\Lc_n^{(1)}+\Lc_n^{(2)} \big)  \psi_{h_3}(z_1,z_2,u_3) = 0 ~, \cr
&& \big(\lb_{-n}^{(3)}-\Lc_n^{(1)}-\Lc_n^{(2)} \big)  \chi_{h_3}(z_1,z_2,\ub_3)=0~,
\eea
leading to
\bea
 \psi_{h_3}(z_1,z_2,u_3)  &=& {z_{12}^{h_3-h_2-h_1} \over (1-z_1 u_3)^{h_3-h_2+h_1}(1-z_2u_3 )^{h_3+h_2-h_1} }~, \cr
 \chi_{h_3}(z_1,z_2,\ub_3)  &=& {z_{12}^{h_3-h_2-h_1} \over (\ub_3-z_1)^{h_3-h_2+h_1}(\ub_3-z_2)^{h_3+h_2-h_1} }~.
 \eea

Correlation functions are obtained by inserting complete sets of states, e.g. for the two-point function
\bea
\langle 0|O_{h}(z_1) O_h(z_2)|0\rangle & = & \int [d^2u] \langle 0| O_h(z_1)|\ub\rangle \langle u|O_h(z_2)|0\rangle \cr
& =&  \int_D [d^2u] \chi_{h}(z_1 ,\ub)\psi_{h}(z_2,u) \cr
& =& (z_1-z_2)^{-2h}~,
\eea
where the integral is computed by series expanding and using (\ref{uint}).   Similarly, the three-point function is
\bea
\langle 0| O_{h_1}(z_1)O_{h_2}(z_2)O_{h_3}(z_3)|0\rangle = \int_D\! [d^2u_3]\chi_{h_3}(z_1,z_2,\ub_3) \psi_{h_3}(z_3,u_3)~.
\eea
Again, the integral is readily performed by series expansion, yielding the standard formula for the three-point function.  For the four-point function, inserting a complete set of states of a given representation $h_p$ yields the corresponding conformal block,
\bea
\langle 0| O_{h_1}(z_1)O_{h_2}(z_2)P_{h_p} O_{h_3}(z_3)O_{h_4}(z_4) |0\rangle   =  \int_D [d^2u_p] \chi_{h_p}(z_1,z_2,\ub_p) \psi_{h_p}(z_3,z_4,u_p)~.
\eea
Taking $z_1=\infty$, $z_2=1$, $z_3=0$, $z_4=z$  (after multiplying by $z_1^{2h_1}$) gives us the standard result for the conformal block in terms of the cross ratio $z$,
\bea
 \langle 0| O_{h_1}(\infty)O_{h_2}(1)P_{h_p} O_{h_3}(0)O_{h_4}(z) |0\rangle  & = & z^{h_p-h_3-h_4} \int_D [d^2u_p] (1-\ub_p)^{h_{12} -h_p}(1-zu_p)^{h_{34}-h_p} \cr
 & = &   z^{h_p-h_3-h_4} {_2}F_1( h_p-h_{12}, h_p-h_{34}, 2h_p ;z)~.
 \eea

Next we turn to the computation of matrix elements of a primary operator between two descendant states.  Since SL(2,$\R$) descendant operators are simply derivatives of primaries, these matrix elements are easily extracted by taking derivatives of the primary three-point function.   The oscillator  formalism is introduced here to set the stage for the Virasoro case, where it is actually useful.

First consider the matrix element
\bea
\langle h_1|  (L_1)^m O_{h_2}(z)(L_{-1})^n  |h_3\rangle = \langle 0|O_{h_1}(\infty) (L_1)^m O_{h_2}(z)(L_{-1})^n O_{h_3}(0)|0\rangle~.
\eea
Inserting complete sets of states we obtain
\bea
\langle h_1|  (L_1)^m O_{h_2}(z)(L_{-1})^n  |h_3\rangle =  \int\! [d^2u_1] [d^2u_3] (\lb^{(u_1)} _{-1})^m\cdot 1~ \Omega(z_2,u_1,\ub_3) (l^{(u_3)} _{-1})^n\cdot 1~,
\eea
where we defined
\bea
\Omega(z_2,u_1,\ub_3)  = \langle u_1| O_{h_2}(z_2)|\ub_3\rangle~.
\eea
Alternatively, let $|u^n\rangle$ correspond to the state whose (unnormalized) wavefunction is $\psi(u)=u^n$.  In this basis, the matrix elements are
\bea\label{umat}
\langle u_1^m |O_{h_2}(z_2)|\ub_3^n\rangle =   \int\! [d^2u_1] [d^2u_3]  u_1^m ~ \Omega(z_2,u_1,\ub_3) \ub_3^n~.
\eea
We determine $\Omega(z_2,u_1,\ub_3)$ by noting that it satisfies the differential equations
\bea  \big( l_n^{(u_1)}-\lb_{-n}^{(u_3)} +\Lc_n^{(z_2)}\big) \Omega(z_2,u_1,\ub_3) =0~,
\eea
which has solution
\bea \Omega(z_2,u_1,\ub_3) = (z_2-\ub_3)^{h_1-h_2-h_3}(1-z_2u_1)^{h_3-h_1-h_2}(1-u_1\ub_3)^{h_2-h_1-h_3}~.
\eea
In the monomial basis, the matrix elements in (\ref{umat}) are obtained by isolating the $\ub_1^n u_3^n$ term in the series expansion of $ \Omega(z_2,u_1,\ub_3)$.   It is straightforward to verify that this approach yields the same matrix elements as obtained by differentiating the primary three-point function.

\subsection{Virasoro}

We now turn to the Virasoro case.   After arriving at an oscillator representation the subsequent steps will proceed as for the SL(2,$\R$) case, although closed form expressions will not be available for all quantities of interest, and so we develop recursion relations that can be implemented numerically.

\subsubsection{Oscillator representation from linear dilaton CFT}

Let $X(z)$ be a free boson with the mode expansion
\bea
 \p X(z) = -i \sum_{m=-\infty}^\infty {\alpha_m \over z^{m+1}}~,
\eea
with
\bea  [\alpha_m,\alpha_n]=m\delta_{m,-n}~.
\eea
The linear dilaton theory has a stress tensor
\bea T(z)=-{1\over 2} \p X \p X + V\p^2 X~,
\eea
and we take $V$ to be real.
The Virasoro generators, obtained as $T(z) = \sum_m {L_m \over z^{m+2}}$, are
\bea\label{virgen}
 L_m &=& {1\over 2}\sum_{n=-\infty}^\infty \alpha_{m-n}\alpha_n +i(m+1)V\alpha_m~,\quad (n\neq 0) \cr
L_0& =& {1\over 2} (\alpha_0)^2 +\sum_{n=1}^\infty \alpha_{-n}\alpha_n +iV\alpha_0~.
\eea
The $L_n$ obey the Virasoro algebra with central charge
\bea c= 1+12 V^2~.
\eea
The adjoint relations $L_n^\dagger =L_{-n}$ are induced from
\bea
\alpha_n^\dagger & = & \alpha_{-n}~,\quad (n\neq 0) \cr
\alpha_0^\dagger & = & \alpha_0+2iV~.
\eea
The $\alpha_0$ relation is understood as resulting from the background charge appearing in the linear dilaton action.  On the sphere the action includes the term ${V\over 4\pi} \int\! d^2 \sigma \sqrt{g}R X = 2V\alpha_0$, so in and out states must have a zero mode momentum differing by this amount in order to obtain a non-vanishing inner product.

We consider representations with zero mode eigenvalue
\bea
\alpha_0 = \sqrt{2} \lambda -iV~,
\eea
where $\lambda$ is real.  We also define $\mu$ via
\bea
\mu = -{V\over \sqrt{2}} = -\sqrt{c-1\over 24}~.
\eea
The $L_0$ eigenvalue of the primary state, denoted  $h$, is then
\bea
h= \lambda^2 + \mu^2~.
\eea

We now represent the nonzero mode oscillators in terms of differential operators acting on holomorphic functions on the plane.   We write
\bea
\alpha_n= {i\over \sqrt{2} } {\p \over \p u_n}~,\quad \alpha_{-n}= -i\sqrt{2} n u_n~,\quad (n>0)~.
\eea
The adjoints $\alpha_n^\dagger = \alpha_{-n}$ are implied by the inner product
\bea
\big(f(U),g(U)\big) =   \int\! [dU] \overline{f(U)}g(U)~.
\eea
with measure
\bea
[dU] =  \prod_{n=1}^\infty d^2u_n {2n\over \pi} e^{-2n  u_n\ub_n}~,
\eea
where the normalization is chosen so that $(1,1)=1$.
We are using the notation $U=\{u_1,u_2, \ldots\}$.   The integration is over the full plane for each $u_n$.

After some rearranging of terms the Virasoro generators (\ref{virgen}), which we now denote by $l_n$,  take the form
\bea\label{vira}
 l_0 & =& h+ \sum_{n=1}^\infty n u_n {\p\over\p u_n}~, \cr
l_k &=& \sum_{n=1}^\infty n u_n {\p\over\p u_{n+k}} -{1\over 4} \sum_{n=1}^{k-1} {\p^2 \over \p u_n \p u_{k-n} }+(\mu k+i\lambda){\p \over \p u_k}~,\quad\quad k>0 \cr
l_{-k} &= &\sum_{n=1}^\infty (n+k) u_{n+k} {\p\over\p u_{n}} - \sum_{n=1}^{k-1}n(k-n) u_n u_{k-n}+2k(\mu k-i\lambda)u_k~,\quad\quad k>0~.
\cr &&
\eea
The middle sums in $l_{\pm k}$ are absent for $k=1$.  These can be verified to obey the Virasoro algebra with $c= 1+24\mu^2$.   The barred generators $\lb_k$ are obtained from the $l_n$ by replacing $u_n\rt \ub_n$ combined with complex conjugation on the  $i$'s.

\subsubsection{Wavefunctions}
\label{app:wavefunctions}

The general strategy is the same as in the SL(2,$\R$) case.  The primary wavefunction is defined as
\bea \psi_h(z,U)=\langle U|O_h(z)|0\rangle~.
\eea
As in (\ref{psieq}) but now using $L_n|0\rangle=0$ for $n\geq -1$,  we derive
\bea (l_n+\Lc_n)\psi_h(z,U)=0~,\quad n \geq -1~.
\eea
It suffices to solve this for $n=-1$ along with the boundary condition $\psi_h(0,U)=1$.  Explicitly, the $n=-1$  equation is
\bea
\p_z \psi_h(z,U)  = \sum_{n=1}^\infty (n+1) u_{n+1} { \p \over \p u_n} \psi_h(z,U) +2(\mu-i\lambda) u_1 \psi_h(z,U)~,
\eea
and the solution is
\bea \psi_h(z,U) = \exp \Big\{{2(\mu-i\lambda) \sum_{n=1}^\infty z^n u_n}\Big\}~.
\eea
This form of this wavefunction is easily understood from the linear dilaton point of view.  A primary operator in the linear dilaton theory is $e^{ip X(z)}$.  Taking $p=\alpha_0$ and using the expressions above, one readily finds $e^{ipX(z)} \cdot 1 = \psi_h(z,U)$, up to an overall normalization factor.

Similarly,
\bea \chi_h(z,\Ub) = \langle 0|O_h(z)|\Ub\rangle
\eea
obeys
\bea  (- \Lc_n + \lb_{-n} ) \chi_h(z,\Ub)~,\quad n\geq -1~,
\eea
and is given by
\bea\label{chiw} \chi_h(z,\Ub) = z^{-2h}\overline{\psi_h(z^{-1},U)} =   z^{-2h}\exp \Big\{{2(\mu+i\lambda) \sum_{n=1}^\infty z^{-n} \ub_n}\Big\}~.
\eea

As in the SL(2,$\R$) case we define
\bea \psi_{h_3}(z_1,z_2,U_3) &=& \langle U_3 |O_{h_1}(z_1)O_{h_2}(z_2)|0\rangle~, \cr
 \chi_{h_3}(z_1,z_2,\Ub_3) &=& \langle 0 |O_{h_1}(z_1)O_{h_2}(z_2)|\Ub_3\rangle~,
\eea
which  obey
\bea\label{psieqs} && \big(l_n^{(3)}+\Lc_n^{(1)}+\Lc_n^{(2)} \big)  \psi_{h_3}(z_1,z_2,U_3) = 0 ~, \cr
&& \big(\lb_{-n}^{(3)}-\Lc_n^{(1)}-\Lc_n^{(2)} \big)  \chi_{h_3}(z_1,z_2,\Ub_3)=0~,\quad  n\geq -1~.
\eea
Unlike the SL(2,$\R$) case, in general no closed form expression for these functions is known.\footnote{The exception is the case $c=1$, $h_1=h_2=h_3=h_4=1/16$ \cite{zamolodchikov1986two}.} It is however possible to solve (\ref{psieqs}) as a series expansion in the oscillator variables.  The four-point Virasoro block is expressed in terms of these functions as
\bea
\langle 0| O_{h_1}(z_1)O_{h_2}(z_2)P_{h_p} O_{h_3}(z_3)O_{h_4}(z_4) |0\rangle  & = & \int [dU] \chi_{h_p}(z_1,z_2,\Ub_p) \psi_{h_p}(z_3,z_4,U_p)~.\cr
&&
\eea

\subsubsection{Matrix elements}

As discussed in the main text, a key virtue of the oscillator construction is that it supplies us with a convenient orthonormal basis.     The inner product between two monomial wavefunctions is
\bea
 \big(  u_1^{m_1} u_2^{m_2} \ldots ,   u_1^{m'_1} u_2^{m'_2} \ldots \big) = \prod_{n=1}^\infty {\Gamma(m_n+1)\over (2n)^{m_n} } \delta_{m_n,m'_n}~.
\eea
Our orthonormal basis elements are then provided by the normalized monomials
 \bea
 \psi_{\{m_a\}}(U) = {  (u_1)^{m_1} (u_2)^{m_2} \ldots \over \sqrt{S_{1,m_1} S_{2,m_2} \ldots}}~,\quad   S_{n,m_n} = {  {\Gamma(m_n+1)\over (2n)^{m_n}}} ~.
 \eea

Given a pair of wavefunctions, $f(U)$ and $g(V)$ associated to representations $h_U$ and $h_V$, the corresponding matrix element of the primary operator $O_{h}(z)$ is
\bea
 \langle f | O_{h}(z)|g\rangle = \int\! [dU][dV] \Omega(z,U,\Vb)\ov{ f(U)} g(V)~, \eea
where
\bea
\Omega(z,U,\Vb) = \langle U|O_{h}(z)|\Vb\rangle~.
\eea
The identity $ [L_n,O_{h}(z)] + O_{h}(z) L_n - L_nO_{h}(z)=0$  implies the relations
\bea\label{omeqs}
 \big(\Lc_n^{(z)}+ l_n^{(U)}-\lb_{-n}^{(\Vb)} \big) \Omega(z,U,\Vb) =0~,\quad  n=0, \pm 1, \pm 2, \ldots ~.
\eea
Note that it is consistent to impose the equations for all $n$ since the commutation relations of $ \Lc_n^{(z)}+ l_n^{(U)}-\lb_{-n}^{(\Vb)}$ are those of a centerless Virasoro algebra.  It follows that if we expand $\Omega(z,U,\Vb)$ terms of normalized monomials the expansion coefficients are the normalized matrix elements of the primary $O_{h}(z)$,
\bea
\Omega(z,U,\Vb)= \sum_{\{m_a\}, \{m'_b\}}  \langle  h_U, \{m_a\} |O_{h_2}(z)| h_V, \{m'_b\} \rangle  \psi_{\{m_a\}}(U)  \psi_{\{m'_b\}} (\Vb)~.
\eea
Our task is therefore to solve (\ref{omeqs}) for $\Omega(z,U,\Vb)$.

\subsubsection{Recursion relations}
\label{app:rec}
The $n=0$ equation in (\ref{omeqs})  is
\bea
 \Big[ -z\p_z +\sum_{n=1}^\infty  n \Big(u_n{\p \over \p u_n} - \vb_n {\p \over \p \vb_n}\Big)  -h +h_U-h_V      \Big] \Omega(z,U,\Vb) =0~.
 \eea
This fixes the $z$ dependence,
\bea
 \Omega(z,U,\Vb) = z^{h_U-h-h_V}F\left(z^{1}u_1, z^{2}u_2, \ldots; z^{-1} \vb_1 ,z^{-2} \vb_2,\ldots \right)~.
 \eea
To fully remove the $z$-dependence consider the action of $\Lc_n$ followed by setting $z=1$.  Since $\Lc_n=-z^{n+1}\p_z-(n+1)hz^n$ we have
\bea
\Lc_n \Omega(z,U,\Vb)\Big|_{z=1}  = \big(-nh -l_0^{(U)}+\lb_0^{(\Vb)}\big) F(U,\Vb)~.
\eea
The system of equations thus becomes
\bea
  \big(-nh -l_0^{(U)}+\lb_0^{(\Vb)} +l_n^{(U)}-\lb_{-n}^{(\Vb)}\big) F(U,\Vb)  =0~,\quad n =\pm 1, \pm 2,\ldots~.
\eea
We now write
\bea
 F(U,V) = \sum_{p,q=0}^\infty F_{p,q}(U,\Vb)
 \eea
where $F_{p,q}(U,\Vb)$ has level $(p,q)$, defined as  $l_0^{(U)} F_{p,q}(U,\Vb)= (h_U+p)F_{p,q}(U,V)$ and $\lb_0^{(\Vb)} F_{p,q}(U,\Vb)$ $= (h_V+q)F_{p,q}(U,\Vb)$.  $l_n^{(U)}$ lowers $p$ by $n$ units, and $\lb_n^{(\Vb)}$ lowers $q$ by $n$ units.   Therefore, the equations at fixed level $(p,q)$ are
\bea
 \big[-nh -l_0^{(U)}+\lb_0^{(\Vb)} \big] F_{p,q} +l_n^{(U)}F_{p+n,q} -\lb_{-n}^{(\Vb)}F_{p,q-n} =0~.
 \eea
So we need to solve this system for all $n\neq 0$ and all $p,q \geq 0$. In fact, we only need to consider the equations $n=-2,-1,0,1,2$, since the the commutation relations imply that the equations with $|n|>2$ will then be automatically satisfied.    We start from $F_{0,0}=1$, and then compute the higher level terms recursively, level by level.  For example, at level $1$ we find
\bea\label{lowsol}
 F_{1,0} &=& { h+h_U-h_V \over \mu+i\lambda_U} u_1 ~,\cr
 F_{0,1}& =& { h-h_U+h_V \over \mu-i\lambda_V} \vb_1~.
 \eea
By examining the transformation of the recursion relations under $U\leftrightarrow V$, $h_U \leftrightarrow h_V$, $p\leftrightarrow q$, $n\leftrightarrow -n$ we see that the solution will obey
\bea  F_{p,q}(U,\Vb;h_U,h_V)= \ov{F_{q,p}(V,\Ub;h_V,h_U)}~,
\eea
as exhibited by (\ref{lowsol}).
%
%

\subsection{Analytic continuation in $h$}
\label{app:h}

In the above we have taken $\lambda$ to be real, which implies $h\geq \mu^2 = {c-1\over 24}$. To access $h< \mu^2$ we cannot simply take $\lambda$   to be imaginary in our existing formulas, since the corresponding $l_n$ would then furnish a non-unitary  representation of Virasoro.   The correct approach corresponds to performing an analytic continuation in $h$ of our final results.  However, since we wish to work with numerical parameter values we need rules to implement this at the level of our equations. We now provide these rules.

In our formulas above, we defined the overbar operation to act on oscillators as $U\rt \overline{U}$ combined with complex conjugation, $i\rt -i$.  Noting that the only place where $i$ appears in our formulas is in the combination $i\lambda$, we can equally well redefine the overbar to act as $U\rt \overline{U}$ combined with $\lambda \rt -\lambda$.    Use of this rule leads to results at $h<\mu^2$ which are the analytic continuation of those from $h> \mu^2$,

In practice this works as follows.  For $h<\mu^2$ we have two possible  imaginary values of $\lambda$, and we write $\lambda^\pm = \pm i \sqrt{\mu^2 -h}$.    We can use $\lambda^+$ in the expressions for the  $l_n$ generators. The $\lb_n$ generators are then obtained by replacing $U\rt \Ub$ and $\lambda^+ \rt \lambda^-$.  We then solve our equations as before.

In (\ref{chiw}) we gave the relation between the $\chi_h$ and $\psi_h$ wavefunctions, whose inner product yields the Virasoro conformal block.  The version of this relation that holds uniformly for all $h\geq 0$ is
\bea
\chi_h(z,\Ub) = z^{-2h} \psi_h(z^{-1},U)\Big|_{U\rt \Ub, \lambda \rt -\lambda}~.
\eea

\subsection{Expectation value of $\langle L_{k}L_{-k}\rangle$}
\label{app:lklk}
We have seen that wavefunctions of the form
$\Psi = u_1^{m_1}u_2^{m_2}u_3^{m_3}\cdots$ form an orthogonal basis. This is a descendant at level $N$
\begin{align}
	l_0^{(U)} \Psi = \left(h_U+\sum_{n=1}^\infty n m_n\right)\Psi = (h_U+N)\Psi~.
\end{align}
$h$ is the conformal dimension of the primary in this section. We would like to evaluate the expectation value $\langle \Psi | l^{(U)}_{k}l^{(U)}_{-k}|\Psi \rangle$.
Let's us proceed first for $k$ being odd. We consider the action of $l^{(U)}_k$ on this state
\begin{align}
	l^{(U)}_{k} \Psi = F_k  \Psi , \qquad F_k \equiv \left[ \sum_{n=1}^\infty n u_n {m_{n+k}\over  u_{n+k}} -{1\over 4} \sum_{n=1}^{k-1} {m_n m_{k-n}\over  u_n   u_{k-n} }+(\mu k+i\lambda_U){m_k \over u_k} \right].
\end{align}
A further action of $l^{(U)}_{-k}$ yields
\begin{align}
	l^{(U)}_{-k} l^{(U)}_{k} \Psi ~=~&\sum_{p=1}^\infty (p+k) u_{p+k} \left({\p F_k\over\p u_{p}} \Psi + F_k {\p \Psi\over\p u_{p}} \right) - \sum_{p=1}^{k-1}p(k-p) u_p u_{k-p}F_k \Psi \nn \\&~+2k(\mu k-i\lambda_U)u_kF_k \Psi .
\end{align}
While computing the inner product the terms which will yield a non-zero contribution are of the kind $(\text{constant})\times \Psi(U)$, the other terms will have a vanishing inner product with the conjugate $\Psi(\bar U)$. The non-vanishing contribution is from
\begin{align}\label{some-terms}
	l^{(U)}_{-k} l^{(U)}_{k} \Psi \ \supset \ &\bigg[\sum_{p=1}^\infty p(p+k)m_{p+k}   +  {\sum_{p=1}^{\infty}p(p+k)m_{p+k}m_p} \nn \\ &+\frac{1}{2} \sum_{p=1}^{k-1}p(k-p) m_p m_{k-p}    +2k(\mu^2 k^2+\lambda_U^2)m_k\bigg]\Psi .
\end{align}
This leads to the expectation value
\begin{align}\label{osc-ev}
	&\vev{l^{(U)}_{-k}l^{(U)}_k}_\Psi \equiv {\vev{\Psi^\dagger l^{(U)}_{-k} l^{(U)}_{k} \Psi  } \over \vev{\Psi^\dagger   \Psi  }  }\nn \\
	&=   \sum_{p=1}^\infty p(p+k)m_{p+k} {(m_p+1)} +\frac{1}{2} \sum_{p=1}^{k-1}p(k-p) m_p m_{k-p}    +(2kh_U+ \tfrac{c-1}{12}(k^3-k))m_k ,
\end{align}
where we used the relations $\lambda_U^2 = h_U-\mu^2 $ and $\mu^2 = (c-1)/24$.

When $k$ is even, we need to be careful about the $p=k/2$ term in the second term of $l_{k>0}$ in \eqref{viro}. The result is
\begin{align}\label{osc-ev-even}
	\vev{l^{(U)}_{-k}l^{(U)}_k}_\Psi  = & \sum_{p=1}^\infty p(p+k)m_{p+k}   {(m_p+1)} + {\frac{k^2}{16} m_{k/2}(m_{k/2}-1) }  +\frac{1}{2} \sum_{p=1}^{k/2-1}p(k-p) m_p m_{k-p}  \nn \\ & + \frac{1}{2} \sum_{p=k/2+1}^{k-1}p(k-p) m_p m_{k-p} +(2kh_U+ \tfrac{c-1}{12}(k^3-k))m_k~.
\end{align}
These expressions can be summed over $k$ and be used to obtain the second KdV charge, $\mathcal{I}^{(2)}(u)=\, \normord{T^2(u)}$.

\subsubsection*{Expectation value in a typical state}
We can use the above analysis to make contact with \cite{Datta:2019jeo} in which the 2-point function of the stress tensor in typical states was studied. Once we write the stress tensors in form of a mode expansion, the key ingredient becomes the expectation value of $\vev{L_{-k}L_k}$ in a typical state.
A typical state is a descendant of level $N$ of the form $\Psi$ here, with the partitions $m_j$ of $N$ distributed according to the Boltzmann distribution. The mean is given by the Bose-Einstein function
\begin{align}
	\vev{m_j} = \left(\frac{1}{e^{\pi j /\sqrt{6N}}-1}\right) ~.
\end{align}
The second moment is given by
\begin{align}\label{second-moment}
	\vev{m_j^2} = \frac{(e^{\pi j /\sqrt{6N}}+1)}{(e^{\pi j /\sqrt{6N}}-1)^2}~.
\end{align}

At large descendant level $N$, we have a large number of typical states and we can replace $m_j$ by its mean.  Although the variance of $m_j$ itself can be large, the variance of the operator product of two stress-tensors, $T(w)T(0)$, can be shown to be small \cite{Datta:2019jeo} and this justifies the replacement by the mean.
We shall return to the case of $k$ even later.  For $k$ being odd, we have
\begin{align}\label{osc-ev-2}
	\vev{l^{(U)}_{-k}l^{(U)}_k}_\Psi  =&   \sum_{p=1}^\infty \frac{p(p+k)}{(e^{\pi (p+k) \over \sqrt{6N}}-1)(1-e^{-\pi p \over \sqrt{6N}})}     +\frac{1}{2} \sum_{p=1}^{k-1}\frac{p(k-p)}{(e^{\pi (k-p) \over \sqrt{6N}}-1)(e^{\pi p \over \sqrt{6N}}-1)}  \nn \\   &+{(2kh_U+ \tfrac{c-1}{12}(k^3-k))\over  e^{\pi k \over \sqrt{6N}}-1}
\end{align}
We can then perform the rescaling $k \to \frac{\sqrt{6N}}{\pi} K,~p \to \frac{\sqrt{6N}}{\pi} P$ and, since we are at large $N$,
we can convert the sums to integrals
\begin{align}\label{osc-ev-3}
	\vev{l^{(U)}_{-k}l^{(U)}_k}_\Psi  \simeq &   \left(\frac{\sqrt{6N}}{\pi}\right)^3\int_{0}^{\infty} dP \frac{P(P+K)}{(e^{P+K}-1)(1-e^{-P})}        \nn \\
	&+\frac{1}{2} \left(\frac{\sqrt{6N}}{\pi}\right)^3\int_{0}^{K} dP\frac{P(K-P)}{(e^{K-P}-1)(e^{P}-1)}  +{(2kh_U+ \tfrac{c-1}{12}(k^3-k))\over  e^{\pi k \over \sqrt{6N}}-1}~.
\end{align}
The integrals appearing above are given by
\begin{align}
	I_1&=  \int_{0}^{\infty} dP \frac{P(P+K)}{(e^{P+K}-1)(1-e^{-P})}  =  \frac{1}{ ( e^K-1)} \left[-K \text{Li}_2\left(e^{-K}\right)-2 \text{Li}_3\left(e^{-K}\right)+\tfrac{\pi ^2 K}{6} +2\zeta(3)  \right],
	\nn \\
	I_2&={1\over 2}\int_{0}^{K} dP\frac{P(K-P)}{(e^{K-P}-1)(e^{P}-1)}  = \frac{1}{ ( e^K-1)}\left[ {-  K \text{Li}_2\left(e^{K}\right)+2  \text{Li}_3\left(e^{K}\right)- \tfrac{\pi ^2 K}{6}-2 \zeta (3)-{ \tfrac{K^3}{12}}}\right]. \nn
\end{align}
Adding the two above we get\footnote{This uses the following identity between polylogs and Bernoulli polynomials
	\begin{align}
		\operatorname{Li}_{n}(z) + (-1)^n \,\operatorname{Li}_{n}(1/z) = -\frac{(2\pi i)^n}{n!} ~B_n \left( \frac{1}{2} + {\log(-z) \over {2\pi i}} \right) \qquad (z ~\not\in ~]0;1])\nn  .
\end{align}}
\begin{align}
	I_1 +I_2
	&= \frac{1}{ ( e^K-1)}\left[\frac{K^3}{12}+\frac{\pi^2 K}{3}\right]~.
\end{align}
%
Putting this back in \eqref{osc-ev-3}, we get, in the regime $k \sim O(N)$ (these give the dominant terms in the thermodynamic limit of the $\vev{T(w)T(0)}$ correlator)
\begin{align}\label{micro-lklk}
	\vev{l^{(U)}_{-k}l^{(U)}_k}_\Psi  \simeq &     { \frac{1}{ ( e^{\pi k \over \sqrt{6N}}-1)} \left[ \frac{k^3}{12}+2Nk\right]  }  +{(2kh_U+ \tfrac{c-1}{12}(k^3-k))\over  e^{\pi k \over \sqrt{6N}}-1} \nn \\
	\approx&   \frac{1}{e^{\pi k \over \sqrt{6N}}-1} \left[2k(h_U +N )  + \tfrac{c}{12}k^3\right]~.
\end{align}
This matches with the thermal expectation value of $L_{-k}L_k$ (see \eqref{avg-1} below). In the last step we got rid of the term $\tfrac{c-1}{12}k$ since it is suppressed compared to the others that add it. This provides an additional verification that typical states reproduce thermal stress tensor correlators.

For the case of even $k$ and $k\sim O(N)$, the additional terms in \eqref{osc-ev-even} are of $O(N)$ while the other terms are $O(N^{3/2})$. This leads to the same result as \eqref{micro-lklk}.

\subsubsection*{$\langle L_{-k} L_{k}  \rangle$ in the canonical ensemble}
We can contrast the above result with the thermal expectation value of $L_{-k}L_k$. As shown in \cite{Maloney:2018hdg,Datta:2019jeo}, this can be obtained from
the thermodynamic limit of $\langle L_{-k} L_{k}  \rangle$ of a single Virasoro module. Here we use, $\tau=i\beta/L$ and $q=e^{2\pi i \tau}$. The conformal dimension of the primary is related to the temperature as $h_U=(c-1)L^2/24\beta^2$
\begin{align}
	\langle L_{-k} L_{k}  \rangle_{\rm m} &= \frac{q^k}{1-q^k} \left[2k(q\pd_q+\tfrac{c}{24})   Z_{h_U}(q) + \tfrac{c}{12}(k^3-k)Z_{h_U}(q)\right]\nn .
\end{align}
Here $Z_{h_U}(q)$ is the character of a primary of dimension $h_U$ and is given by  $q^{h_U-(c-1)/24}\eta(q)^{-1}$. Therefore
\begin{align}
	\langle L_{-k} L_{k}  \rangle_{\rm m}
	&= \frac{q^k}{1-q^k} \left[2k(h_U -  \tfrac{1}{24} E_2(q) )  + \tfrac{c}{12}k^3\right]Z_{h_U}(q)~.
\end{align}
No approximations have been made so far and this is an exact result. We would like to evaluate this for $q=e^{-2\pi\beta/L}$ in the limit $L\to \infty$. The S-modular transformation of the Eisenstein series, $E_2(q)$, can be used to obtain the behaviour in this regime and we get
\begin{align}\label{avg-1}
	\langle L_{-k} L_{k}  \rangle_{\rm m}
	\approx    \frac{1}{ ( e^{-{2\pi\beta k \over L}}-1)} \left[2k(h_U +\tfrac{L^2}{24\beta^2} )  + \tfrac{c}{12}k^3\right]Z_{h_U}(q)~.
\end{align}
Note that this is the unnormalized expectation value and hence the additional factor of $Z_{h_U}$. As expected, this matches \eqref{micro-lklk} upon using the usual relation of the descendant level of a typical state to the temperature, $N=L^2/24\beta^2$.

\bibliographystyle{bibstyle2017}
\bibliography{collection}

\end{document}